\documentclass[sigconf,10pt]{acmartmod}
\usepackage{graphicx}
\usepackage{balance}
\usepackage{amsmath}
\usepackage[inline]{enumitem}
\usepackage{subcaption}
\usepackage{array}
\usepackage{algorithm}
\usepackage{algpseudocode}
\usepackage{xfrac}
\usepackage{wrapfig}
\usepackage{oubraces}
\usepackage{xspace}
\usepackage{listings} 
\usepackage{caption}
\usepackage{color}
\usepackage[utf8]{inputenc}
\usepackage{tikz}
\usetikzlibrary{patterns}
\usetikzlibrary{arrows}	

\DeclareMathOperator*{\argmin}{\arg\!\min}
\DeclareMathOperator*{\abs}{\textrm{abs}}
\DeclareMathOperator*{\sign}{\textrm{sign}}

\algblockdefx[UponEvent]{Upon}{EndUpon}[2][event]{\textbf{Fully merged sample arrives} $\langle$\vspace{0mm}\emph{#2}$\rangle$ \textbf{do}}{\textbf{end}}

\newcommand{\usolid}[2]{%
    \tikz[baseline=(todotted.base)]{
        \node[inner sep=1pt,outer sep=0pt] (todotted) {#2};
        \draw[solid,color=#1] (todotted.south west) -- (todotted.south east);
    }%
}%
\newcommand{\udot}[2]{%
    \tikz[baseline=(todotted.base)]{
        \node[inner sep=1pt,outer sep=0pt] (todotted) {#2};
        \draw[dotted,color=#1] (todotted.south west) -- (todotted.south east);
    }%
}%
\newcommand{\udash}[2]{%
    \tikz[baseline=(todotted.base)]{
        \node[inner sep=1pt,outer sep=0pt] (todotted) {#2};
        \draw[dashed,color=#1] (todotted.south west) -- (todotted.south east);
    }%
}%
\newcommand*\circled[1]{\tikz[baseline=(char.base)]{
            \node[shape=circle,draw,inner sep=0.5pt,fill=white] (char) {#1};}}

\newcommand{\trustClockSymbol}{\raisebox{-0.8mm}{\includegraphics[trim=17px 17px 15px 15px, clip, height=12px]{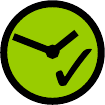}}\xspace}
\newcommand{\noTrustClockSymbol}{\raisebox{-0.8mm}{\includegraphics[trim=17px 17px 15px 17px, clip, height=12px]{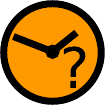}}\xspace}

\newcommand{\para}[1]{\noindent\textbf{#1.}}
\renewcommand{\paragraph}[1]{\noindent\textbf{#1.}}

\makeatletter
\renewcommand\subsubsection{\@startsection{subsubsection}{3}{\z@}%
                                     {1mm}%
                                     {1mm}%
                                     {\normalfont\large\itshape}}%
\let\oldsubsubsection\subsubsection
\renewcommand{\subsubsection}[1]{\oldsubsubsection{#1}\par\noindent\ignorespaces}
\makeatother

\setcopyright{none}
\acmArticle{4}
\hypersetup{draft}

\begin{document}

\title{SENSE: Scalable Data Acquisition from Distributed Sensors with Guaranteed Time Coherence}

\author{Jonas Traub}
\affiliation{\institution{Technische Universität Berlin}}
\author{Julius Hülsmann}
\affiliation{\institution{Technische Universität Berlin}}
\author{Sebastian Breß}
\affiliation{\institution{Technische Universität Berlin}}
\author{Tilmann Rabl}
\affiliation{\institution{HPI Potsdam}}
\author{Volker Markl}
\affiliation{\institution{Technische Universität Berlin}}

\begin{abstract}
Data analysis in the Internet of Things (IoT) requires us to combine event streams from a huge amount of sensors.
This combination (join) of events is usually based on the time\-stamps associated with the events.
We address \mbox{two challenges} in environments which acquire and \mbox{join events in the IoT:}

First, due to the growing number of sensors, we are fa\-cing the performance limits of central joins with respect to throughput, latency, and network utilization.
Second, in the IoT, diverse sensor nodes are operated by different organizations and use different time synchronization techniques.
Thus, events with the same timestamps are not necessarily recorded at the exact same time and joined data tuples have an unknown time incoherence.
This can cause undetected failures, such as false correlations and wrong predictions.

We present SENSE, a system for scalable data acquisition from distributed sensors.
SENSE introduces time coherence measures as a fundamental data characteristic in addition to common time synchronization techniques.
The \textit{time coherence} of a data tuple is the time span in which all values contained in the tuple have been read from sensors.
We explore concepts and algorithms to quantify and optimize time coherence
and show that SENSE scales to thousands of sensors, operates efficiently under latency and coherence constraints, and adapts to changing network conditions.
\end{abstract}

\maketitle

\section{Introduction}\label{sec:chp3-introduction}
Stream analysis systems have access to a growing number of data streams from sensor nodes in the Internet of Things (IoT)~\cite{GartnerIoT}.
Analytical applications correlate these data streams to facilitate fast event detection and respective reactions.
Typically, sensor measurements consist of a timestamp $t$ and a value $v$.
Stream analysis systems~\cite{carbone2015apache,toshniwal2014storm,zaharia2016apache} and time series databases~\cite{bader2017survey,turnbull2018monitoring} join measurements from distributed sensors to tuples in the form $\langle t,v_1,\dots,v_n\rangle$.
This central correlation has three major scalability issues:
\begin{enumerate*}
\item it results in a vast amount of parallel network connections at a central system,
\item it misses edge computing opportunities, and
\item it relies on costly stream joins.
\end{enumerate*}

Moreover, applications assume that joint sensor data tuples represent a concise snapshot of all values (i.e., measurements) taken at time $t$.
Imprecise timestamps can cause application failures such as wrong correlations, missed event detections, and false predictions.
For example, think about a traffic accident:
Multiple sensors record videos, measure speed, capture braking time, and report safety distances.
When joining sensor measures, it is crucial to ensure time coherence to decide if a driver's reaction time was too high, or if the safety distance was too small \textit{before} one car braked.

However, in practice, joint tuples are affected by an unknown \textit{time incoherency}:
In the IoT, many different users and organizations operate diverse sensor nodes such as smart phones, weather stations, smart watches, and connected cars~\cite{lee2015internet}.
The precision of system clocks on these devices depends on various aspects including time synchronization, clock drift, security vulnerabilities, and intended manipulation.
Most devices synchronize their clocks through the Network Time Protocol (NTP) \cite{ntpv2,ntpv3,ntpv4}, which has known security issues~\cite{dowling2016authenticated,liveoverflow2017time}:
For example, the \textit{NIST Authenticated NTP Service} requires to ship keys via postal mail for authentication~\cite{NISTNTP}.
NTPv4 provides an authentication method based on public keys~\cite{ntpv4}.
However, this method is rarely used and vulnerable to brute force attacks as it uses small 32-bit seeds~\cite{dowling2016authenticated}.
In practice, most sensor nodes do not use NTP authentication and are vulnerable to spoofing attacks~\cite{wang2015time}.
In addition, each device requires an oscillator to track the progress of time.
Many sensor nodes use cheap quartz crystals~\cite{raouf2013minimize}, rely on the processor frequency~\cite{bergsma2013howaccurate}, or monitor external oscillations~\cite{tauber2018electric,meyer2018how,bytycy2018serbia}. This causes widespread clock drifts.
Last but not least, many applications give an incentive to deliberately manipulate the system time~\cite{liveoverflow2017time}.
For example, a parcel service can delay clocks on handset scanners to hide late deliveries and a software user can delay clocks to hide the expiration of a license.

We conclude that timestamps require an input validation~\cite{scholte2012have,liveoverflow2017time} and that we need a system which allows for \textit{scalable data gathering with guaranteed time coherence}.

We present SENSE, a system which addresses the problem of \textit{time incoherence} and solves \textit{scalability challenges} in the Internet of Things. 
SENSE combines \textit{central stream joins} with a new architecture based on \textit{sensing loops} to ensure scalability.
The system utilizes edge computing capabilities to avoid central computation and transmission bottlenecks.
The time coherence management of SENSE is a complementary addition to time synchronization techniques. Advances in time synchronization~\cite{lasassmeh2010time, sichitiu2003simple, sivrikaya2004time} reduce clock deviations, which leads to an improved estimated coherence in SENSE.
The estimated coherence of a tuple is the difference between sensor read times assuming correct clock synchronization among clocks of sensor nodes.

We take into account that sophisticated techniques for time synchronization have not yet seen widespread adoption in the IoT and that timestamps may be manipulated.
To this end, we allow for specifying an incoherence limit for tuples which is maintained by the system. 
For each tuple, SENSE provides a guaranteed time coherence which is independent of clock synchronization.
Coherence estimates and coherence guarantees then work as key data characteristic in applications and quantify the result precision.
We further introduce algorithms which optimize the resource utilization under latency and time coherence constraints.
Finally, we add fault-tolerance mechanisms to make our system robust against sensor and network failures.

\vspace{2mm}
\noindent Overall, this paper makes the following contributions:
\begin{enumerate}[topsep=4pt,itemsep=4pt]
	\item We present an architecture for acquiring values from large numbers of sensors with guaranteed time coherence and low latency (Section~\ref{sec:architecture}).
	\item We introduce time coherence as a fundamental data characteristic of sensor data tuples (Section~\ref{sec:CgandCe}).
	\item We optimize the coherence of result tuples and provide coherence guarantees which are independent of clock synchronization among sensor nodes (Section~\ref{sec:tolerances}).
	\item We pre-schedule sensor reads for future requests (Section~\ref{sec:scheduling-techniques}) and provide fault-tolerance mechanisms for sensor and network failures (Section~\ref{sec:failure-handling}).
	\item We experimentally evaluate our solution and show that it scales to thousands of sensors, provides low latencies, and operates efficiently (Section~\ref{sec:evaluation}).
\end{enumerate}
\vspace{1.5mm}

\noindent
In the remainder of this paper,
we present an application example in Section~\ref{sec:intro-example} 
and discuss sources of incoherence in Section~\ref{sec:sources-of-incoherence}.
We then
present our contributions and experiments in Sections~\ref{sec:architecture}-\ref{sec:evaluation}.
Finally, we discuss related work in Section~\ref{sec:related-work}
and conclude in Section~\ref{sec:conclusion}.

\section{Application Example:\newline Precision of Multilateration}\label{sec:intro-example}
\begin{figure}[t]
\includegraphics[]{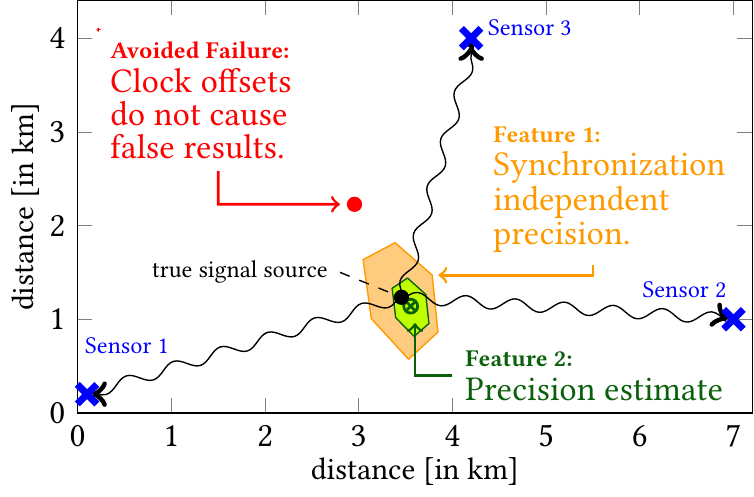}
\caption{Multilateration example.}\label{fig:intro-example}
\end{figure}

In this section, we introduce time coherence failures and show how the coherence measures of SENSE prevent these failures in an example application.

\paragraph{Application Example}
We demonstrate SENSE in a multilateration application.
Multilateration is a common technique to locate the source of a signal by measuring the times when the signal arrives at three or more sensors.
Common applications of multilateration are finding the epicenter of an earthquake, locating positions of lightnings in thunderstorms, or locating aircrafts.
We chose this example, because it allows for nicely visualizing the use of coherency measures with a small number of sensors. In practice, application usually correlate data from larger number of sensors.

The delta among the arrival times of a signal at different sensors exposes the delta in the distances between the signal source and sensors.
From these distances, multilateration applications compute the position of the signal source (e.g., an aircraft or a lightning).
We visualize our example in Figure~\ref{fig:intro-example}.
We use three sensors (blue crosses) to locate the source of thunder (i.e., the positions of lightnings) in a thunderstorm.
The signal disseminates from its source (black point) with sonic speed which is about 343$\frac{m}{s}$ (767mph).
We know the positions of sensors and compute the positions of lightnings.
We provide more details about our calculations, which are not required to follow the example, in Appendix~\ref{sec:trilateration-calc}.

\paragraph{Possible Failures}
Offsets among sensor node clocks add an observation error to the arrival times of a signal,
which results in false locations.
For example, the {\color{red}\usolid{red}{red}} point in Figure~\ref{fig:intro-example} shows the computed lightning location which results from Sensor 2 being two seconds behind Sensor~1 and Sensor~3 being two seconds ahead of Sensor~1.
Existing techniques simply assume that sensors report correct arrival times. Thus, they neither detect, quantify, nor prevent this error.

\paragraph{Feature 1 - Synchronization Independent Precision}
\definecolor{cgarea}{HTML}{FF9900}%
\definecolor{darkgreen}{HTML}{0B610B}%
\linebreak In SENSE, users can specify a precision requirement which is \textit{independent of clock synchronization} among sensor nodes.
This precision requirement is reflected in an upper limit for coherence guarantees called $C_{g_{max}}$.
For example, the {\color{cgarea}\usolid{cgarea}{orange}} area in Figure~\ref{fig:intro-example} depicts the precision for $C_{g_{max}}$=2 seconds.
With $C_{g_{max}}$=2 seconds, SENSE will locate the signal in the orange area even if sensor node clocks have arbitrary offsets.
Thus, SENSE would avoid the failure in Figure~\ref{fig:intro-example}  and locate the signal source in the orange area instead.

\paragraph{Feature 2 - Precision Estimate}
The coherence estimate allows for computing the area in which we expect the signal source assuming perfect synchronization among sensor node clocks%
({\color{darkgreen}\usolid{darkgreen}{green}} area in Figure~\ref{fig:intro-example}).
In our example, sensors have a read time precision of $\pm$0.5 seconds, which is reflected in $C_e$.
The result is a slight deviation in the location of the signal source (green point).
The strength of our solution is the ability to quantify possible deviations (green area) instead of returning calculated locations only.

In summary, SENSE avoids failures, by providing precision guarantees which are independent of clock synchronization, and by quantifying precision estimates.

\section{Sources of Incoherence}\label{sec:sources-of-incoherence}
There exist many reasons for deviations among sensor node clocks which impact time coherence. In this section, we highlight important observations with respect to these sources of incoherence.
These observations lead to system requirements which motivate design decisions in SENSE.

\begin{figure}[t]
\captionsetup[subfigure]{justification=centering,position=t}
\begin{subfigure}[t]{0.35\linewidth}%
  \includegraphics[]{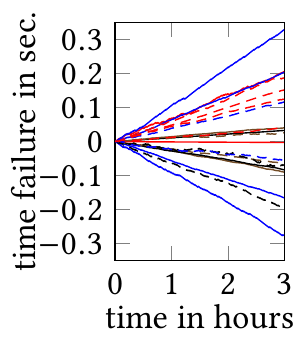}
  \caption{System Clock on Raspberry Pi~\cite{bergsma2013howaccurate}}\label{fig:clock-drift-a}
\end{subfigure}%
\hspace{-1mm}%
\begin{subfigure}[t]{0.35\linewidth}%
  \includegraphics[]{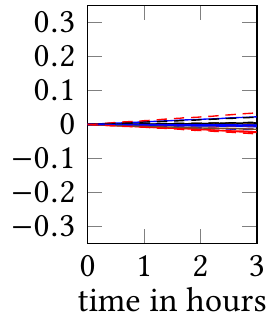}
  \caption{Ras Clock RTC (14\$/pc.)~\cite{nxp2014PCF2127}}\label{fig:clock-drift-b}
\end{subfigure}%
\hspace{-3.3mm}%
\begin{subfigure}[t]{0.35\linewidth}%
  \includegraphics[]{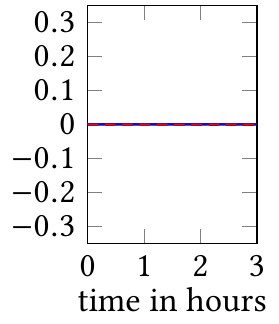}
  \caption{High precision clock (23\$/pc.)~\cite{iqd2017PCF2127}}\label{fig:clock-drift-c}
\end{subfigure}%
	\vspace{-4mm}
	\caption{Clock drifts of 20 simulated clock instances for three different clock types.}\label{fig:clock-drifts}
\end{figure}

\vspace{2mm}
\paragraph{Clock Drift}
In Figure~\ref{fig:clock-drifts}, we visualize clock drifts of three different clock types:
\begin{enumerate*}[label=(\alph*)]
\item Raspberry Pi system clocks which use the processor frequency as reference~\cite{bergsma2013howaccurate},
\item Real time hardware clocks which have an integrated quartz crystal and cost about 14\$ per unit~\cite{nxp2014PCF2127}, and
\item high precision clocks which cost about 23\$ per unit~\cite{iqd2017PCF2127}.
\end{enumerate*}
We provide supplementary technical information
for all clocks and our simulation in Appendix~\ref{sec:clocks-for-clock-drift}.
\noindent We make three observation in Figure~\ref{fig:clock-drifts}:

\noindent\textbf{1)}\textbf{ Cheap clocks are heavily affected by clock drift.}
\noindent
Regular system clocks on Raspberry Pis, which are widely used as sensor nodes, drift up to $\pm$0.14s/hour (3.36s/day).
Remembering our example in Section~\ref{sec:intro-example}, this drift causes major errors in the result after just a few hours uptime.

In addition, many devices do not have an own oscillator and depend on external oscillations which could possibly fail, such as the oscillation of the electricity grid.
For example, in March 2018 thousands of devices, including electric meters and inverters of solar plants, accumulated more than 6 min. delay,
due to a conflict between Kosovo and Serbia about who is responsible for providing additional energy~\cite{tauber2018electric}.
This caused the oscillation to stay below the desired 50Hz for more than a month~\cite{meyer2018how,bytycy2018serbia}.
Hence, a conflict between two countries affected clocks everywhere in Europe.
Although the power grid was brought back to its regular frequency~\cite{geuss2018european}, one still finds wrong timestamps in the logs of many smart meters and solar plants.
This example shows that clocks may malfunction for surprising and unexpected reasons.

\noindent\textbf{2)}\textbf{ Sensor nodes with precise clocks are expensive.}%
\noindent We want to correlate data from thousands of sensors in the Internet of Things~\cite{GartnerIoT}.
These sensors are regularly hosted on cheap devices (sensor nodes) such as micro controllers, Arduinos, or Raspberry Pis.
A precise clock alone costs 14\$ or more, whereas a whole Raspberry Pi costs about 35\$ (without a precise clock).
Thus, large numbers of sensor nodes can hardly be upgraded with precise clocks due to the high price.

\noindent\textbf{3)}\textbf{ Cheap clocks measure short time spans precisely.}
Even cheap clocks can measure short time spans precisely.
We call the timeframe between the current time and the read time of a sensor value \textit{age of the value}.
Clock drift is the only factor which pollutes the measured ages of values.
For example, the Raspberry Pi system clock drifts at most 0.006s/min.
We regularly request values from sensors which are only a few seconds old and, thus, use the age of values as a reliable and precise measure on sensor nodes.

\vspace{2mm}
\paragraph{Clock Synchronization}
Sensor nodes limit the impact of clock drifts with repeated clock synchronization.
NTPv4 is one of the most common synchronizations protocols and claims to be precise within \textit{a few tens of milliseconds}~\cite{ntpv4}.
However, when it comes to large numbers of diverse sensors, we cannot rely on precise synchronization among all of them:

\noindent\textbf{1) The precision depends on the network.}
\noindent NTPv4 specifies its precision for \textit{fast local area networks}~\cite{ntpv4}.
Other networks may have larger latencies or more volatile transmission times which leads to more frequent and less reliable re-synchronization.
Especially devices connected via mobile networks may experience precision issues.

\noindent\textbf{2) Not all devices have synchronized clocks.}
\noindent Many low-cost sensor nodes do not synchronize their clocks at all.
For example, many micro controllers just start their clocks when powered on,
only providing a simple up-time counter instead of a real clock.
Since we regularly require the age of values instead of read times, our solution integrates sensors without synchronized clocks seamlessly.

\vspace{2mm}
\paragraph{Resulting System Requirements} We conclude that SENSE shall quantify the uncertainty about times provided by sensor nodes with coherence measures.
SENSE shall
provide synchronization independent coherence guarantees, and limit incoherence with time coherence optimization.

\begin{figure}
\centering%
\captionsetup[subfigure]{justification=centering}%
\begin{subfigure}[b]{0.62\linewidth}
  \centering
  \includegraphics[width=\linewidth]{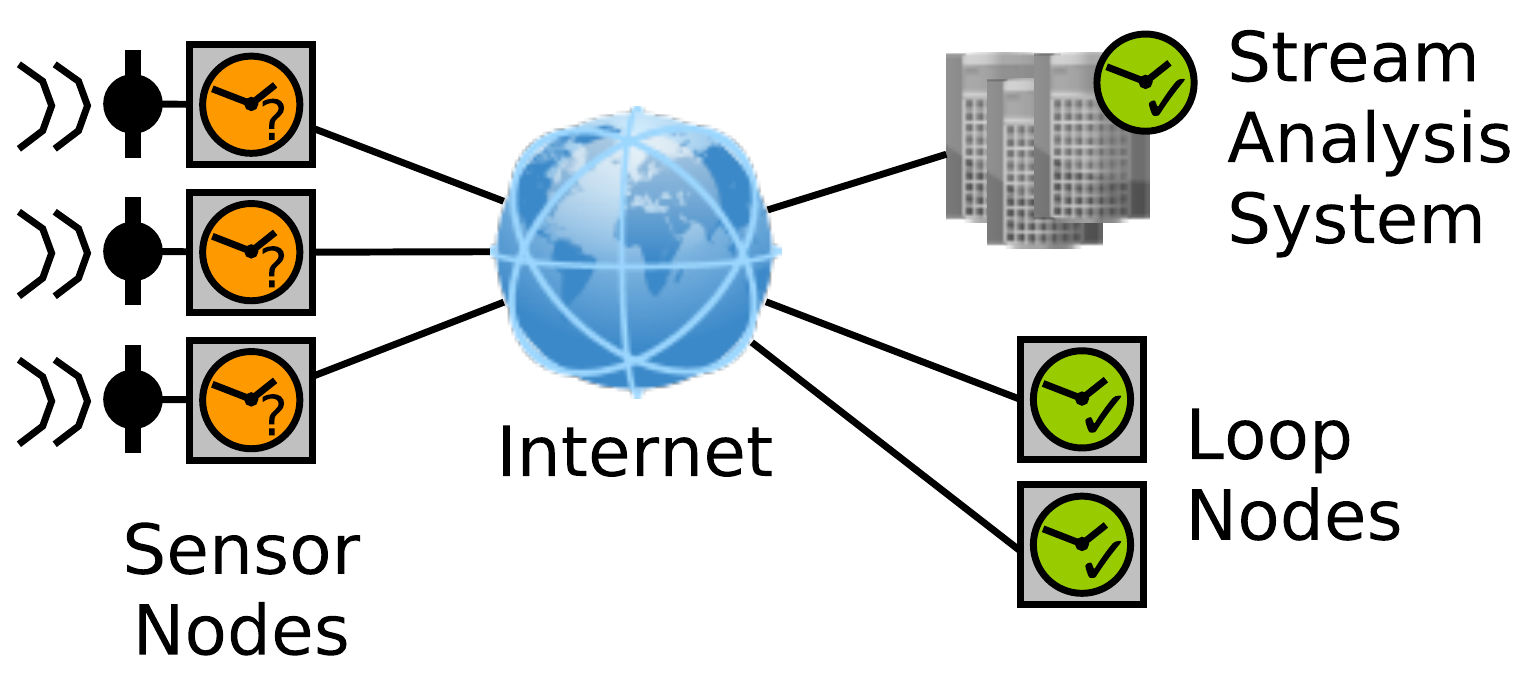}
  \caption{Network connections and locations of clocks.}\label{fig:general-setup-nodes-and-clocks}
\end{subfigure}
\begin{subfigure}[b]{0.36\linewidth}
  \centering
  \includegraphics[width=\linewidth]{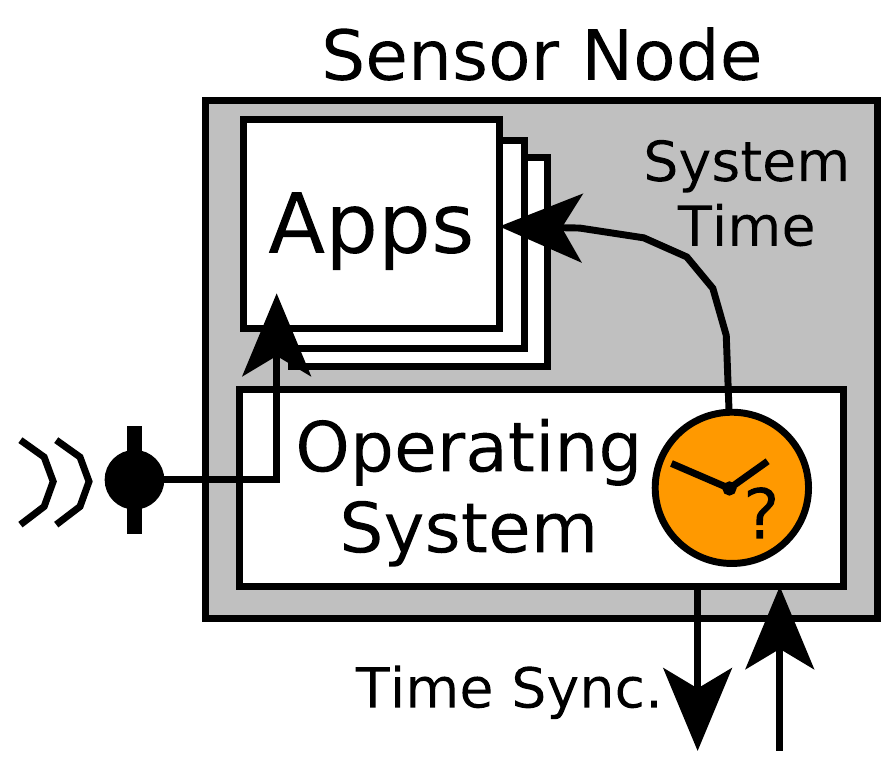}
  \caption{Sensor~node \mbox{architecture}.}\label{fig:general-setup-sensor-node}
\end{subfigure}%
\caption{General network and node setup in the IoT.}\label{fig:general-setup}
\end{figure}

\section{SENSE Architecture}\label{sec:architecture}

In this section we define coherence measures (Section~\ref{sec:coherency-definition}) and introduce the SENSE architecture.
We discuss the \textit{general network and node setup} in Section~\ref{sec:general-iot-setup},
present the \textit{global architecture} which covers the connection and interaction among distributed sensor nodes in Section~\ref{sec:global-arch-overview}, and
describe the \textit{Internal Architecture}, i.e., the software components of sensor nodes, in Section~\ref{sec:internal-architecture}.

\subsection{Definition of Coherence Measures}\label{sec:coherency-definition}

\paragraph{Coherence of a Tuple / $\boldsymbol{C_{real}}$}
Let a tuple contain the values
$v_1,...,v_N$ from $N$ sensors
and let $t_1,...,t_N$ be the times when $v_1,...,v_N$ were measured.
The coherence of a tuple is the timeframe between min($t_1,\dots,t_N$) and max($t_1,\dots,t_N$) in which $v_1,...,v_N$ were measured.

Tuple coherence is a key indicator to detect incoherent tuples and to prevent false analysis results. %
Ideally, all values of a tuple would be measured at the same time, which would result in the optimal coherence 0.

\paragraph{Coherence Estimate of a Tuple / $\boldsymbol{C_e}$}
We use the term \textit{estimate} for the computed coherence ($C_e$) to emphasize the uncertainty about its correctness.
$C_e$ equals $C_{real}$ if there is no offset among sensor node clocks.
In practice $t_1,\dots,t_N$ are obtained from different sensor node clocks and may be affected by clock drift.
Thus, the computed coherence estimate $C_e$ of a tuple may diverge from the true coherence.

\paragraph{Coherence Guarantee of a Tuple / $\boldsymbol{C_g}$}
The coherence guarantee of a tuple is a timeframe which is guaranteed to be larger or equal than the real coherence of the tuple ($C_g \geq C_{real}$).
We use the term \textit{guarantee} to emphasize that $C_{real}$ is guaranteed to be smaller or equal than $C_{g}$ even if there are arbitrary offsets among sensor node clocks.
The guarantee $C_g$ ensures that time coherence is bounded. %

\paragraph{Read Time Deviation / $\Delta_t$}
Each sensor data tuple has a timestamp $t$ which is the desired read time (i.e., request time) for values contained in the tuple.
$\Delta_t$ is the maximum deviation between the desired read time $t$ and any actual read time $t_i$ of a value $v_i$ contained in the tuple.
For example, if we request a tuple with $t$=$5$ and sensor provide values read at $t_1$=$4$ and $t_2$=$7$, then $\Delta_t$=$\max(\abs(5-4),\abs(5-7))$=$2$.

\subsection{General Network and Node Setup}\label{sec:general-iot-setup}
In this section, we discuss our assumptions with respect to network connections, communication protocols, geographic proximity, processing capabilities, and reliable clocks.

Our general setup covers key components of different IoT architectures~\cite{fremantle2014reference, krco2014designing} (Figure~\ref{fig:general-setup-nodes-and-clocks}).
Typically, sensor nodes communicate through the internet and/or local networks via standard TCP/IP protocols~\cite{forouzan2002tcp}.
They may use additional protocols and message brokers to manage their communication such as MQTT~\cite{stanford2013mqtt} or ZeroMQ~\cite{hintjens2013zeromq}.
We require acknowledgements of receipt, but do not make any additional assumptions with respect to communication protocols or connection types (e.g., WiFi, LTE, or cable).

\begin{figure}
  \centering
  \includegraphics[width=0.9\linewidth]{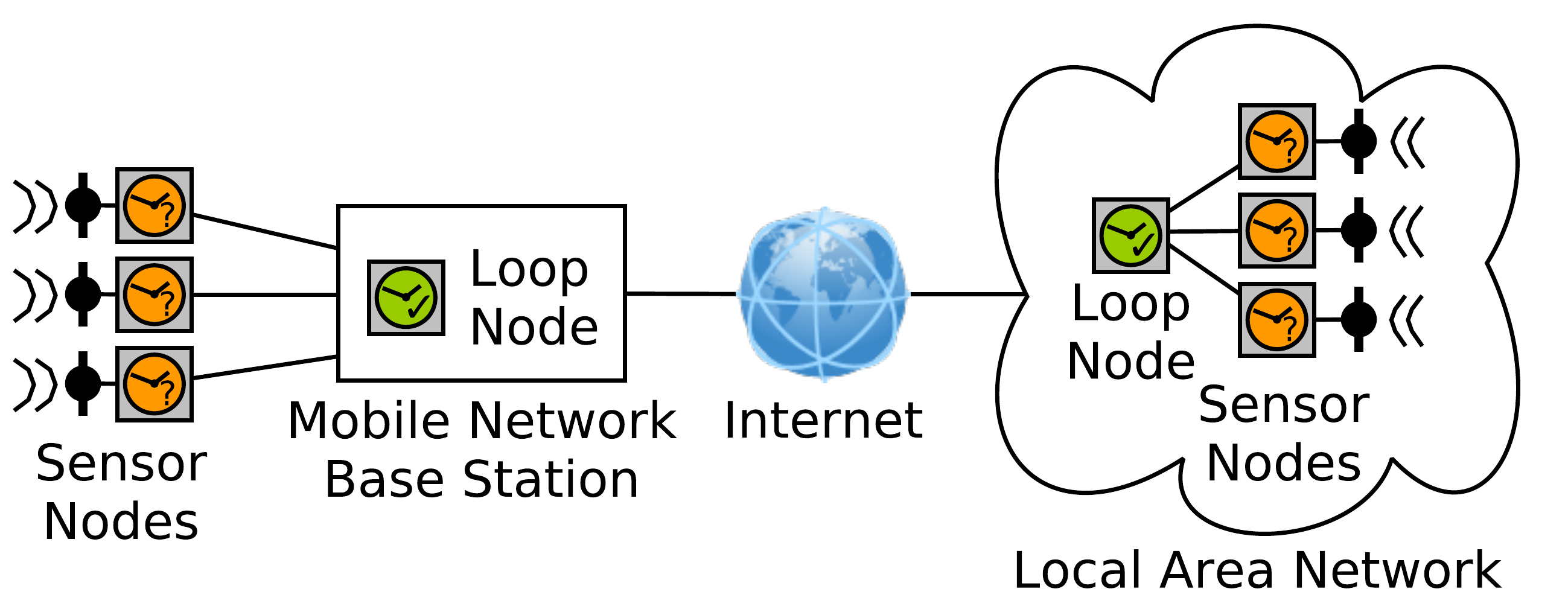}
  \caption{Two examples for exploiting local proximity of loop nodes and sensor nodes.}\label{fig:general-setup-ln-location}
\end{figure}

We differentiate trusted (\trustClockSymbol) and untrusted (\noTrustClockSymbol) clocks.
For \textit{trusted clocks}, we control time synchronization and can validate that their system time is correct.
For \textit{untrusted clocks}, we cannot enforce correctness.
We operate a small number of nodes with trusted clocks which we call \textit{Loop Nodes}. 
Loop nodes are servers which control the data acquisition from sensor nodes (to be discussed in Section~\ref{sec:global-arch-overview}).
Since the number of loop nodes is small, we can afford to equip them with high precision clocks and make sure that trusted clocks are in sync.
The majority of devices are sensor nodes with untrusted clocks.  
Common sensor nodes are smart phones (e.g., Android or iOS systems) and small computers (e.g., Arduinos and Raspberry Pis), which are operated by diverse users and organizations.
Since we do not control the clocks on these devices, we cannot enforce their correctness.

Figure~\ref{fig:general-setup-sensor-node} shows a sketch of the software architecture of sensor nodes.
They consist of an operating system and an application layer.
The operating systems allows for accessing sensors through driver modules~\cite{Giouroukis2019resense} and maintains the system clock.
The application layer (user space) hosts applications which access sensors and system time through the operating system.
On sensor nodes, SENSE runs on the application layer~--~either as an application on its own, or as a library integrated in a host application such as a smart phone app.
Ideally, sensor nodes can schedule sensor reads, have sufficient memory to buffer recent sensor values, and can retrieve these values upon request.
Most sensor nodes fulfill these requirements.
However, we discuss in Section~\ref{sec:internal-architecture} how we integrate sensor nodes which do not have sufficient memory and processing capabilities. 

SENSE exploits geographic proximity if it leads to fast network connections among sensor nodes and loop nodes.
Figure~\ref{fig:general-setup-ln-location} shows two examples of a desirable proximity.
On the left, a loop node is collocated with the base station of a mobile network and manages connected sensor nodes.
On the right, a loop node and sensor nodes are collocated in the same local area network.
Both setups lead to low latency and low jitter\footnote{Network jitter is the variance of transmission times in a network.}, which improves coherence guarantees.
Despite these benefits, SENSE does not require local proximity and adapts to the observed latency, time coherence, and jitter.

\subsection{Global Architecture}\label{sec:global-arch-overview}

\begin{figure}[t]
\captionsetup[subfigure]{justification=centering}
\centering
\setlength{\tabcolsep}{0pt}
\begin{tabular}{m{0.4\hsize}m{0.6\hsize}}
\begin{subfigure}[b]{\linewidth}
  \centering
  \vspace{4mm}
  \includegraphics[width=\linewidth]{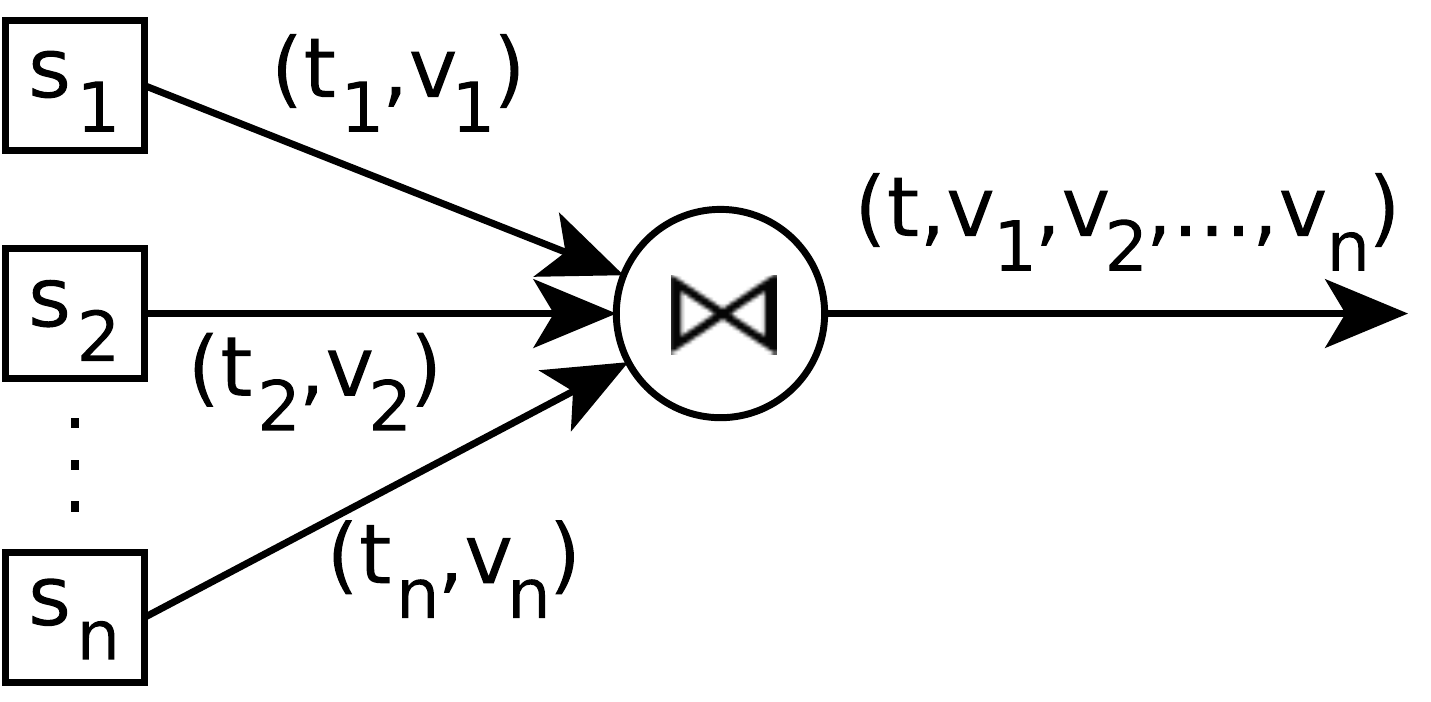}
\caption{Central Join.}\label{fig:central-join}
\end{subfigure}
&
\begin{subfigure}[b]{\linewidth}
  \centering
  \includegraphics[width=\linewidth]{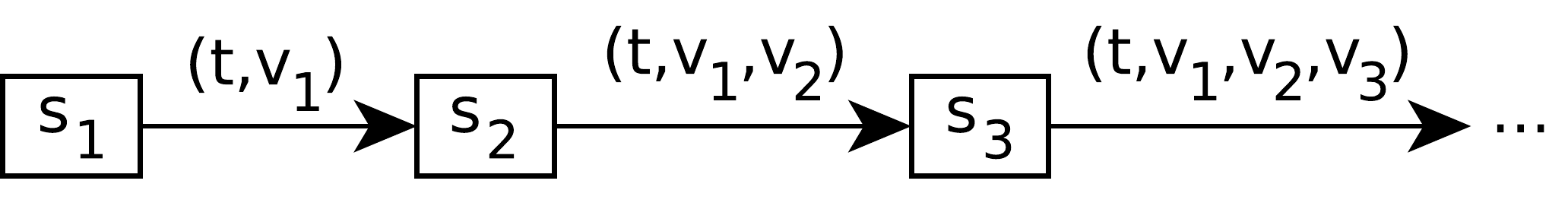}
  \caption{Sensing Pipeline.}\label{fig:sensing-pipeline}
  \vspace{2mm}
\end{subfigure}
\begin{subfigure}[b]{\linewidth}
  \centering
  \includegraphics[width=\linewidth]{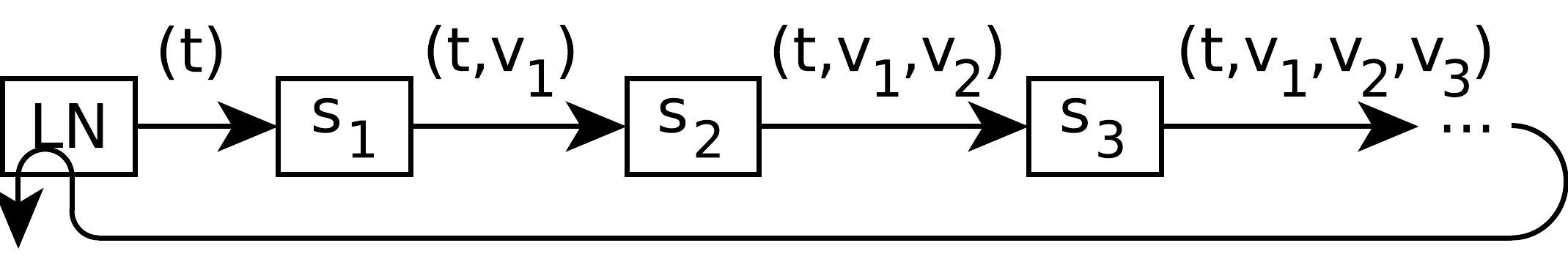}
  \caption{Sensing Loop.}\label{fig:sensing-loop}
\end{subfigure}
\end{tabular}
\vspace{-2mm}
\caption{Topologies for sensor data acquisition.}\label{fig:architechtural-alternatives}
\end{figure}

Two common topologies for sensor data acquisition are \textit{Central Joins} (Figure~\ref{fig:central-join}) and \textit{Sensing Pipelines} (Figure~\ref{fig:sensing-pipeline}).
First, we discuss advantages, disadvantages, and limitations of these topologies.
Then, we introduce \textit{Sensing Loops} (Figure~\ref{fig:sensing-loop}) to overcome the limitations observed before.

In a \textbf{central join} topology, each sensor streams pairs of timestamps ($t$) and values ($v$) to a central server which joins time-value-pairs from all sensors.
Examples are smart phone apps which report values to a server and stream joins in systems such as Apache Flink~\cite{carbone2015apache} and Spark~\cite{zaharia2016apache}.
The main advantage of this solution is its low latency.
The latency is low, because sensors transmit their values directly to the central server.
However, there are two major disadvantages in a central join topology:
\begin{enumerate*}
	\item The join cannot provide any guarantee for the time coherence of result tuples because it relies on the correctness of the timestamps transmitted from sensor nodes.
	\item The solution does not scale to large numbers of sensors because of a limited number of concurrent network connections and an increasing join complexity.
\end{enumerate*}

In a \textbf{sensing pipeline}, one node initiates a request and passes it to the succeeding node in the pipeline.
Each node in the pipeline adds a value from its sensor and forwards the tuple to the next sensor until the tuple contains all values.
Example applications are Sensor Networks~\cite{akyildiz2002survey, demers2003cougar,madden2005tinydb,vidyasagar2009wireless}.
Users submit queries to the network which collects data from sensors - possibly performs in-network computation~\cite{madden2002tag, sharaf2003tina} - and finally returns result tuples to the user.
The advantage of sensing pipelines is that they overcome scalability limitations of central servers which receive result tuples.
Since the result tuple is produced in-network, there is only one connection to the receiver instead of a large number of concurrent connection from individual sensors.
Moreover, there is no need to perform a join centrally, which reduces complexity. %

However, we still face latency issues for large numbers of sensors, because each transmission hop adds latency.
Similar to central joins, it is impossible to ensure time coherence, because we rely on separate sensor node clocks.

We overcome the limitations of Central Joins and Sensing Pipelines by extending Sensing Pipelines to \textit{Sensing Loops} and by combining \textit{Sensing Loops with Central Joins}:

\begin{figure}[t]
\centering
\includegraphics[width=0.95\linewidth]{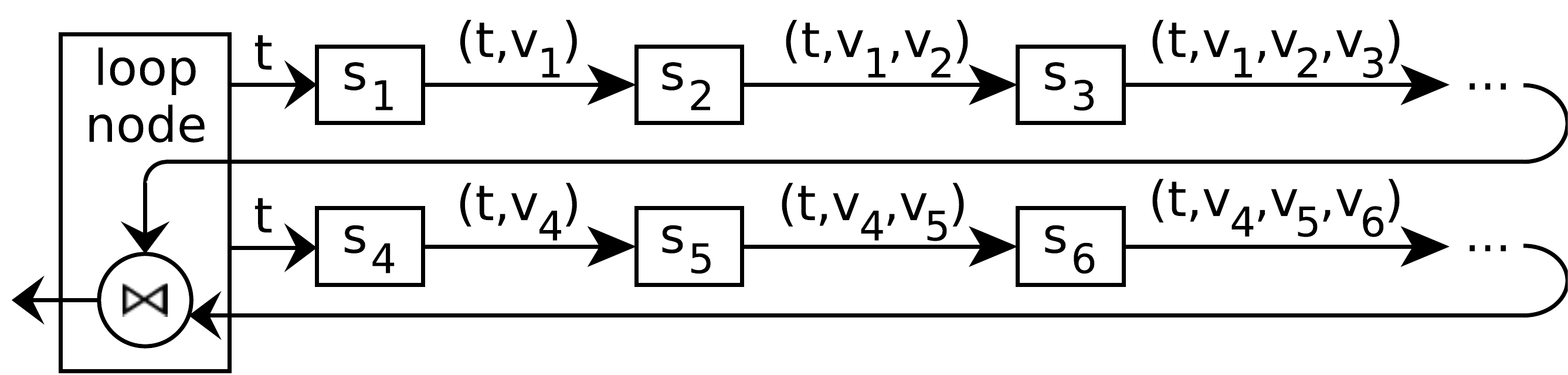}
\caption{Example of an architecture combination.}\label{fig:architectural-combinations}
\end{figure}

\textbf{Sensing loops} (Figure~\ref{fig:sensing-loop}) extend \textit{Sensing Pipelines} with an additional \textit{Loop Node} (LN).
The loop node initiates requests by sending a request timestamp to the first sensor node in the pipeline.
Once the request passes the pipeline, the last sensor node returns the result tuple to the loop node.

\textbf{Enabling Coherence Guarantees.}
Sensing Loops enable coherence guarantees because sensor data tuples pass by the same clock (the loop node clock) twice.
The loop node saves the time a request was sent (start time $l_s$) and observes the time when it receives the result tuple (end time $l_e$).
If the sensors read ad-hoc upon request, all individual values are guaranteed to be gathered within the time interval $ \mathcal I := [l_s, l_e]$, whose diameter $C_g := l_e - l_s$ thus serves as tight upper bound to the actual tuple incoherence, hence adheres to the definition of coherence guarantee.
In the remainder of this chapter, we generalize this intuition by providing a general representation of the time interval $\mathcal I$ for more sophisticated read scheduling approaches. 
The interval's diameter $C_g$ is independent of clock synchronization, as it solely utilizes time distance measurements at the loop node.

\textbf{Scaling to Large Numbers of Sensors.}
A scalability issue of long sensing pipelines results from high latency in too long pipelines.
A scalability issue of central joins results from too many parallel network connections and an increasing computation effort for large numbers of sensors.
We combine both solutions in Figure~\ref{fig:architectural-combinations} to overcome scalability limitations by allowing for multiple sensing loops $p$ which we automatically split and merge depending on latency and coherence requirements interacting with the observed network conditions.
The loop node then centrally joins the results of all sensing loops and subsequently publishes the result. 
For example, a sensing loop can collect values from 100 sensor nodes with less than three seconds latency.
The loop node can join results returned by 100 loops with less than a second latency.
In combination, we can provide a time coherent snapshot from 10000 sensors with less than four seconds latency and a coherence guarantee below three seconds.

We compute a joint $C_g$ at the loop node as follows:
For each pipeline $p$, we 
compute intervals $\mathcal I^{(p)}$ from $t_s$ and $t_e^{(p)}$\hspace{-1mm} as described in the previous paragraph.
Values gathered from nodes that belong to an arbitrary pipeline $p$ are guaranteed to be gathered within the time interval $\mathcal I^{(p)}$, hence all values of the resulting tuple are therefore guaranteed to be gathered during the union of all pipeline-specific intervals $$ \mathcal I := \bigcup\limits_p \mathcal I^{(p)}.$$

\subsection{Internal Architecture}\label{sec:internal-architecture}
\begin{figure}[t]
\centering
\includegraphics[width=\linewidth]{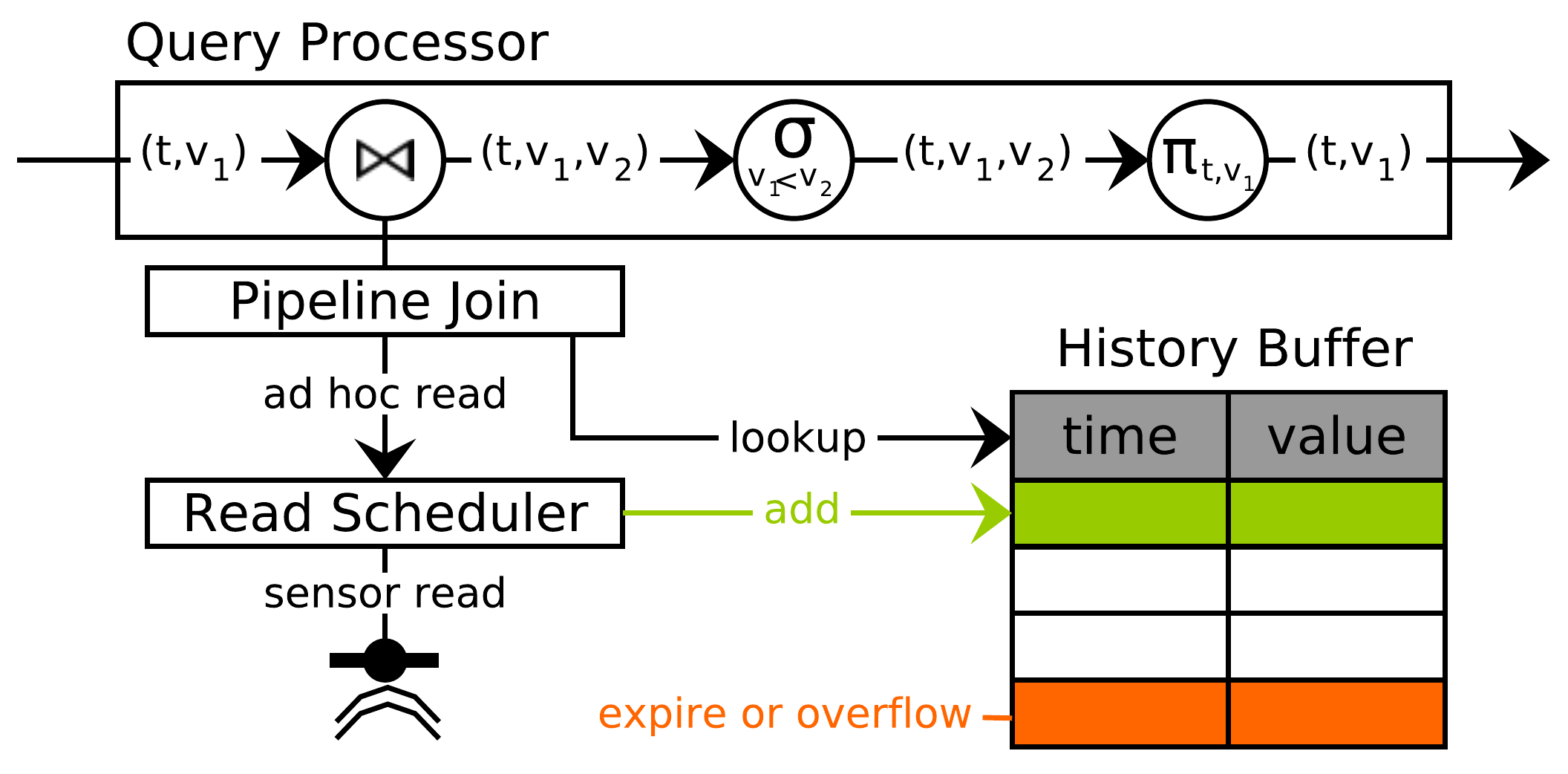}
\caption{Overview of sensor node internals.}\label{fig:sensor-node-internals}
\end{figure}

In the following paragraphs, we describe different sensor node components of the SENSE architecture (Figure~\ref{fig:sensor-node-internals}).

\paragraph{Query Processor}
Our query processor is a stream processor which adopts a tuple-at-a-time processing model similar to common streaming systems such as Apache Flink and Storm.
Each tuple passes through a processing pipeline which includes gathering sensor values (Pipeline Join) as well as data manipulations (e.g., selections or projections)~\cite{madden2005tinydb}.
In this paper, we focus on the time coherence of result tuples with respect to sensor read times.
Accordingly, we focus on data gathering operations in the remainder of the paper.

\paragraph{Pipeline Join}
Our pipeline join is a sophisticated replacement of a simple ad-hoc sensor read.
The pipeline join and its integration are key contributions of this paper.
Instead of reading sensor values ad-hoc when processing a tuple, we join input tuples with a recent sensor measurement.
For each tuple, each sensor node selects the best value from its history buffer with respect to coherence measures.

\paragraph{Read Scheduler}
The read scheduler is an active component which performs sensor reads and adds the resulting values and read timestamps to the buffer of the sensor node.
The read scheduler also performs requested ad-hoc reads.
We use a read scheduler we introduced in our previous work~\cite{traub2017ondemand}. Our scheduler supports reading periodically, scheduling sensor reads at specific times, reading ad-hoc, and diverse adaptive sampling techniques such as AdaM~\cite{trihinas2015adam}, FAST~\cite{fan2012real}, and L-SIP~\cite{gaura2013edge}.
We discuss relevant scheduling approaches for this paper in Section~\ref{sec:scheduling-techniques}.

\paragraph{History Buffer}
The history buffer of a sensor node stores read times and sensor measurements.
The read scheduler adds entries to the buffer.
The history buffer handles the expiration of buffer entries.
Entries expire regularly after a certain time or after they were joined with a tuple.
In case of failures, there may be buffer overflows.
We discuss our expiration and overflow mechanisms in detail in Section~\ref{sec:bufferManager}.

\paragraph{Sensor Node Limitations}
We are aware that there exist sensor nodes which cannot support all components described above and ensure that SENSE seamlessly integrates such nodes.
If a sensor node cannot run a sophisticated read scheduler, we read values from sensors periodically, which is commonly supported by sensor nodes (Section~\ref{sec:scheduling-techniques}).
If a sensor node has a buffer that is too small, we address buffer overflows with an overflow mechanism (Section~\ref{sec:bufferManager}).
If a sensor node cannot process custom operations going beyond sensor reads, we will not assign additional operators to the node~\cite{madden2005tinydb}.
We will also not assign operators to a node which has insufficient computation power.
We observed that our pipeline join requires much less computation effort than performing sensor reads and networking tasks and did not experience any performance issues caused by pipeline joins.

\section{Coherence Guarantees and Coherence Estimates}\label{sec:CgandCe}
In this section, we discuss how SENSE calculates coherence guarantees $C_g$ and coherence estimates $C_e$ for tuples.
We illustrate our discussion with diagrams which are inspired by UML sequence diagrams.
In these diagrams time processes from top to bottom and columns separate locations (i.e., sensor nodes) at which actions take place.

\subsection{Coherence Guarantees}
The loop node observes the coherence guarantee for each sensor data tuple which passes the loop.
We compute the coherence guarantee at the loop node with Formula \ref{eq:Cg}.
\begin{align}
C_g=(l_e-\min(\alpha_1,\dots,\alpha_n))-(l_s-\max(\alpha_1,\dots,\alpha_n))
\label{eq:Cg}
\end{align}
We name the start time of a tuple passing through the loop $l_s$ and the end time $l_e$.
When performing a pipeline join, each sensor $s_i$ provides the age $\alpha_i$ of its value $v_i$ used for the join.
Thus, $\alpha_i$ is the age of $v_i$ at the join time at $s_i$.
$\alpha$ is highly precise for recent measurements, even with cheap clocks, because short time spans are barely affected by clock drifts.

\begin{figure}[t]
\centering
\includegraphics[width=0.68\linewidth]{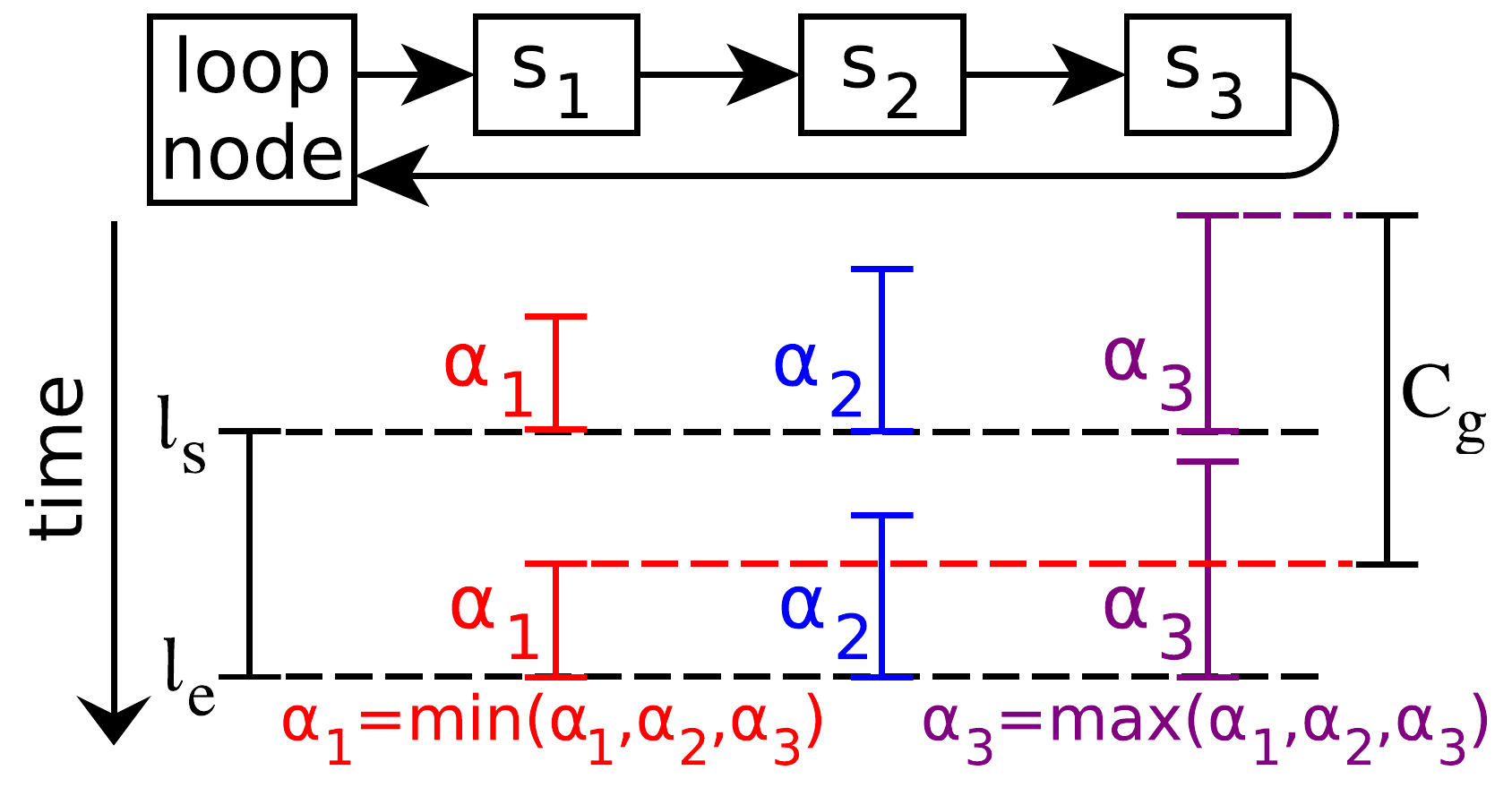}
\caption{Illustration of the coherence guarantee $\boldsymbol C_g$.}\label{fig:Cg}
\end{figure}

Figure~\ref{fig:Cg} illustrates the observation of the coherence guarantee.
We know that all pipeline joins are performed between $l_s$ and $l_e$.
Thus, the latest possible read time of a value in the tuple is $l_e-\text{min}(\alpha_1,\alpha_2,\alpha_3)$, which is $l_e-\alpha_1$ in our example.
The earliest possible read time of a value in the tuple is $l_s-\text{max}(\alpha_1,\alpha_2,\alpha_3)$, which is $l_e-\alpha_3$ in our example.
The coherence guarantee is the timeframe between the earliest and the latest possible read time calculated in Formula~\ref{eq:Cg}.

\subsection{Coherence Estimate}

\begin{figure}[t]
\centering
\includegraphics[width=0.68\linewidth]{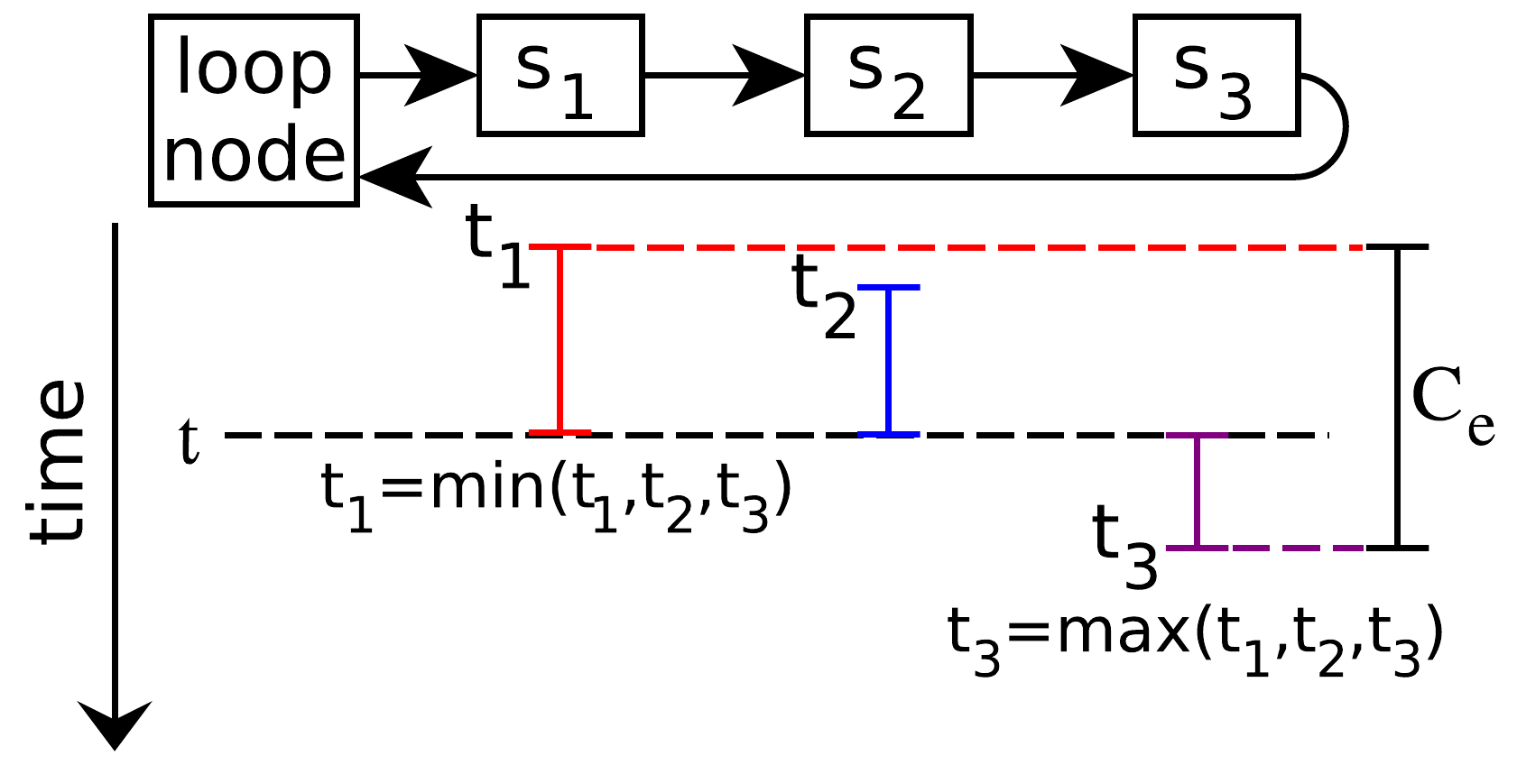}
\caption{Illustration of the coherence estimate $\boldsymbol C_e$.}\label{fig:Ce}
\end{figure}

We calculate $C_e$ at the loop node with Formula~\ref{eg:Ce}.
Figure~\ref{fig:Ce} illustrates the calculation.
Each sensor node $s_i$ provides the read time $t_i$ of its value $v_i$ used for the sensor data tuple.
We take the latest read time of any value (according to sensor node clocks) which is $\max(t_1,\dots,t_n)$ and subtract the earliest read time of any value which is $\min(t_1,\dots,t_n)$.
\begin{align}
C_e=\max(t_1,\dots,t_n)-\min(t_1,\dots,t_n)
\label{eg:Ce}
\end{align}
Formula~\ref{eg:Ce} calculates $C_e$ under the assumption that all sensor node clocks are in sync.
Thus, it provides an estimate which disregards clock deviations.
Alternatively, we can calculate $C_e$ independent of clock synchronization if we know or estimate data transmission latencies instead.
$C_e$ may be more precise based on transmission latencies than based on clock synchronization.
For example, if transmission times are constant (hardware buses), well predictable (dedicated connection), or negligible (fibre optics).

\begin{figure}[t]
\centering
\includegraphics[width=0.69\linewidth]{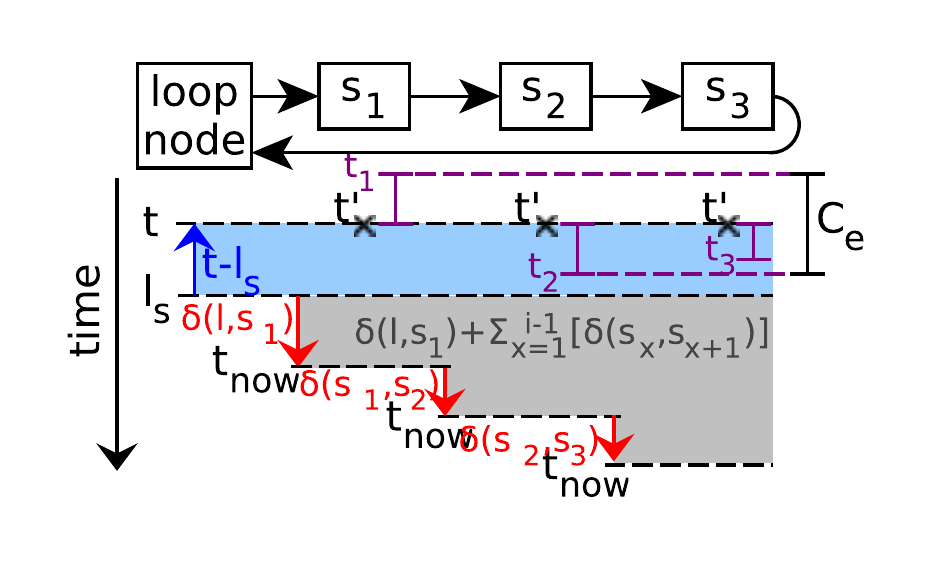}
\caption{$\boldsymbol C_e$ based on transmission times.}\label{fig:CeLatency}
\end{figure}

Figure~\ref{fig:CeLatency} illustrates how we compute $C_e$ based on transmission and processing latencies among sensor nodes.
Let $\delta(s_a,s_b)$ be the transmission and processing latency between two sensors $s_a$ and $s_b$,
and let $\delta(l,s_a)$ be the transmission and processing latency between the loop node and a sensor $s_a$.
Let $t_{now}$ be the current time according to the unsynchronized sensor clock.
Then the desired read time $t'$ according to the unsynchronized sensor clock is given by Formula~\ref{eg:tstrich}.
We perform the pipeline join based on the calculated $t'$ - ignoring the desired read time $t$ contained in the tuple.
Thus, we make the join independent of time synchronization.
\begin{align}
t'&=t_{now}+(t-l_s)-(\delta(l,s_1)+\delta(s_1,s_2)+\dots+\delta(s_{i-1},s_i))\nonumber\\
t'&=t_{now}+(t-l_s)-(\delta(l,s_1)+\sum_{x=1}^{i-1}\delta(s_x,s_{x+1}))
\label{eg:tstrich}
\end{align}
As a result of the join, we receive $v_i$ and $t'_i$.
$t'_i$ is the read time of $v_i$ according to the unsynchronized clock at $s_i$.
We compute $t_i$ from $t'_i$ based on the shift between $t$ and $t'$ with the formula \mbox{$t_i = t'_i+(t-t')$}.
Finally, we can compute $C_e$ from $t_i,\dots,t_N$ as before with Formula~\ref{eg:Ce}.

\subsection{Coherence Tradeoff ($\boldsymbol C_e$-$\boldsymbol C_g$-Tradeoff)}
\label{sec:coherence-tradeoff}

\begin{figure}[t]
\centering
\captionsetup[subfigure]{justification=centering,position=t}
\hspace{-1mm}%
\begin{subfigure}[t]{0.5\linewidth}%
  \centering
  \includegraphics[width=\linewidth]{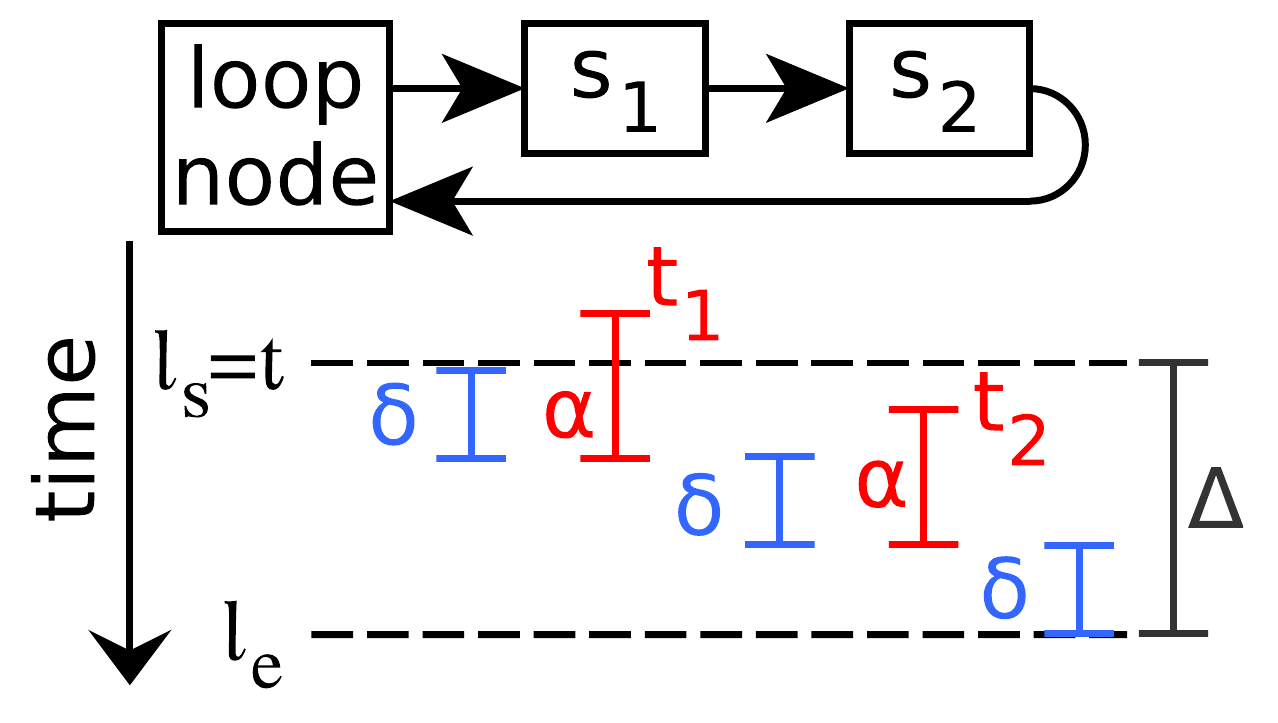}
  $\delta\!=\!2$; $\alpha_{min}\!=\!\alpha_{max}\!=\!3$\\ $C_g\!=\!\Delta$; $C_e\!=\!2$
  \caption{Best possible $\boldsymbol C_g$}
  \label{fig:tradoff-ce-cg-1}
\end{subfigure}%
\hspace{1mm}%
\begin{subfigure}[t]{0.5\linewidth}%
  \centering
  \includegraphics[width=\linewidth]{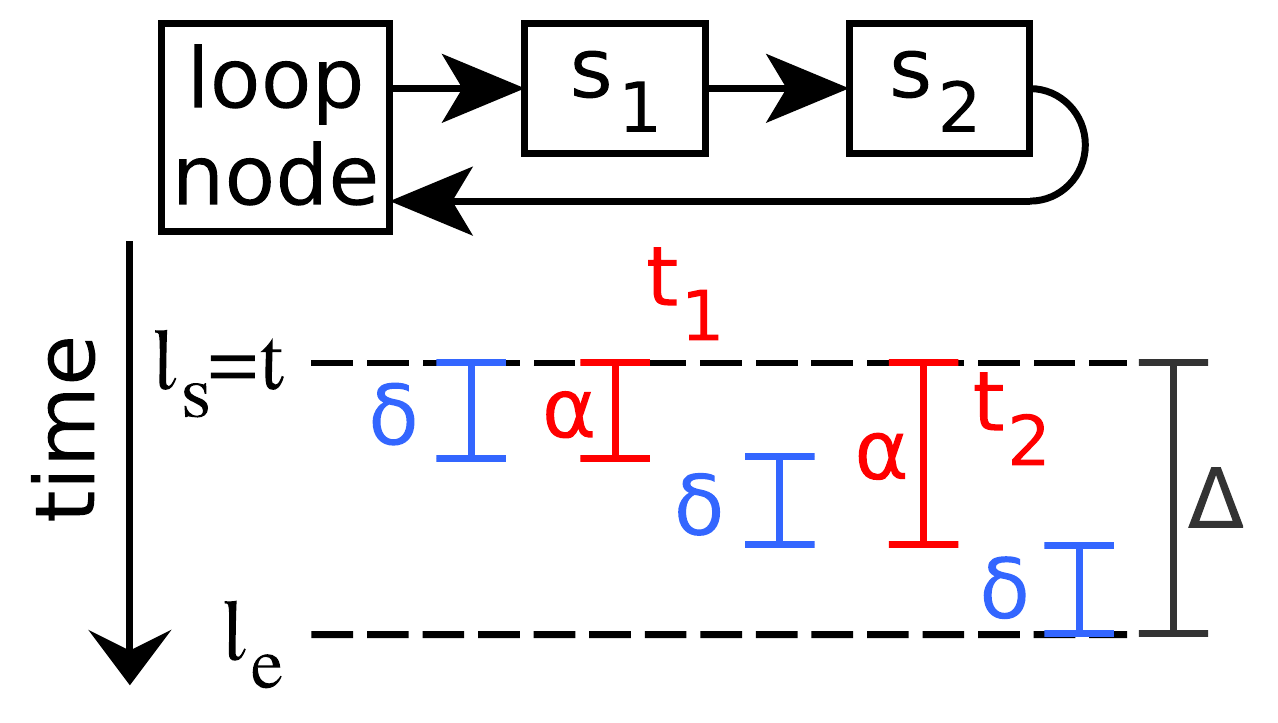}
  $\delta\!=\!2$ ; $t\!=\!t_{min}\!=\!t_{max}$\\
  $C_g\!=\!\Delta+2$ ; $C_e\!=\!0$
  \caption{Best possible $\boldsymbol C_e$}
  \label{fig:tradoff-ce-cg-2}
\end{subfigure}%
\caption{The tradeoff between $\boldsymbol C_e$ and $\boldsymbol C_g$.}
\label{fig:tradoff-ce-cg}
\end{figure}

We compute a coherence guarantee ($C_g$) and a coherence estimate ($C_e$) for each tuple.
There exists a tradeoff between the best possible $C_g$ and the best possible $C_e$ of a tuple.
We show this tradeoff in Figure~\ref{fig:tradoff-ce-cg}. In our example, we specify the mean hop time $\delta$=$2$ and acquire values from $N$=$2$ nodes.
In Figure~\ref{fig:tradoff-ce-cg-1}, we show the optimal read times with respect to $C_g$.
$C_g$ is optimal if $\alpha_{min}$=$\alpha_{max}$.
However, if $\alpha_{min}$=$\alpha_{max}$ (i.e., if all ages are identical), then $t_1,\dots,t_N$ deviate from the request time $t$ which increases $C_e$.
In Figure~\ref{fig:tradoff-ce-cg-2}, we set $t$=$t_{min}$=$t_{max}$ which is the optimum with respect to $C_e$.
In exchange, $\alpha_{max}-\alpha_{min}$ is $2$ which increases $C_g$ respectively.

Typically, users want to specify a precision requirement which translates to an upper limit for $C_g$ which we call $C_{g_{max}}$.
At the same time, users want to receive a result which is as precise as possible.
Usually, a smaller $C_e$ improves the result quality assuming that sensor nodes try to provide correct timestamps for their values.
In the next section, we describe how we tradeoff between $C_g$ and $C_e$ automatically such that we minimize $C_e$ ensuring $C_g \leq C_{g_{max}}$.

\section{Optimizing Time Coherence}\label{sec:tolerances}
In this section, we present how SENSE optimizes the coherence estimate while enforcing an upper bound for the coherence guarantee.

\subsection{Solution Overview}
We automatically adjust two parameters ($\alpha$ and $\mu$) to trade off between the best coherence guarantee and the best coherence estimate.
The loop node computes the coherence guarantee ($C_g$), the coherence estimate ($C_e$), and the roundtrip time ($\Delta$) for each tuple which passes through the sensing loop.
Using these values ($C_g$, $C_e$, and $\Delta$), the loop node continuously tunes $\alpha$ and $\mu$ and disseminates updates to all sensor nodes in the loop.
We distribute $\alpha$ and $\mu$ by attaching them to regular sensor data tuples to prevent additional messages.

When a sensor node receives a tuple (i.e., a sensor data request), it adds a value from its buffer based on the current values of $\alpha$ and $\mu$ (pipeline join).
Let $t$ be the desired timestamp of the tuple and $t_{now}$ be the current time according to the sensor node clock.
To achieve an optimal $C_g$, we would select a value read at $t_{now}$-$\alpha$.
To achieve an optimal $C_e$, we would select a value read at $t$.
Thus, we select a value which was read between $t_{now}$-$\alpha$ and $t$.
The current value of $\mu$ specifies which value between $t_{now}$-$\alpha$ and $t$ we select. In the next sections, we discuss how sensor nodes select values from  buffers based on $\alpha$ and $\mu$ and how loop nodes tune $\alpha$ and $\mu$.

\subsection{Sensor Node Algorithm}\label{sec:sensor-node-algo}
We now discuss how sensor nodes, which are part of a sensing loop, join input tuples (i.e., sensor data requests) with sensor values from their buffers.
A sensor node, which is the $i$-th node in the loop, receives tuples in the form $\langle t,\alpha,\mu,\alpha_{min}, \alpha_{max},t_{min},t_{max},v_1,~\dots~, v_{i-1} \rangle$. The node joins each tuple with a value $\langle t_i,v_i \rangle$ from its local sensor stored in the history buffer.
The resulting output tuple has the format $\langle t,\alpha,\mu,\alpha_{min}, \alpha_{max},t_{min},t_{max},v_1,~\dots~, v_{i} \rangle$.
$\alpha_{min}$ and $\alpha_{max}$ are the minimum and maximum age of any value ($v_1,\dots, v_n$) at the join time at the respective sensor node.
$t_{min}$ and $t_{max}$ are the minimum and maximum timestamps $t_i$ of any value ($v_1,\dots, v_n$) according to sensor node clocks.

We define the optimization function which selects the best available value $v_i$ in Formula~\ref{eq:optFunc}.
This optimization function expresses the tradeoff between $C_e$ (first part) and $C_g$ (second part).
The first part increases linearly for worse $C_e$.
The second part has a higher order to emphasize its weight strongly when approaching an upper bound for $C_g$.
The parameter $\mu \in \mathbb{R}^+$ weights the second part against the first part.%
\begin{align}
\text{opt}(t,\!t_{now},\!\mu,\!\alpha)\!=\!\argmin\limits_{t_i \in \text{Buffer}}[\underbrace{\abs(t_i\!-\!t)}_{\text{cost of }C_e}\!+\!\underbrace{(t_{now}\!-\!t_i\!-\!\alpha)^2}_{\text{cost of }C_g}\!\cdot\mu]\label{eq:optFunc}
\end{align}
It is beneficial to not only base Function~\ref{eq:optFunc} on the desired tuple timestamp $t$, but to also make use of the impact of previously joined values on the coherence tradeoff. 

This adaptation to the optimization function is described in Appendix~\ref{sec:opt-at-node}.

\subsection{Loop Node Algorithm}\label{sec:loop-node-algo}
We now discuss how we tune $\mu$ and $\alpha$ on loop nodes. In addition to variables defined before, our formulas and algorithms in this section use the following variables:
\begin{itemize}[align=parleft]
  \item[$l_s$] Start time of the loop (outbound transmission time).
  \item[$\Delta$] Round trip time for the current tuple. 
  \item[$\delta$] Mean hop time between nodes.
  \item[$N$]  Number of sensor nodes in the sensing loop.
  \item[$D_{max}~$]  Desired upper bound for $C_g$ (max. incoherence).
  \item[$t_l$] Last time $\mu$ and $\alpha$ have been updated.
  \item[$p$] Indicates whether $D_{max}$ was met. Initialized as $0$.
  \item[$s$] Step width exponent (scales step width for $\mu$ updates). Initialized as $0$.
  \item[$w$] Step width for $\mu$ update.
\end{itemize}
Before we discuss how we select $\alpha$ and $\mu$, we define $D_{max}$.
$D_{max}$ is a system internal variable which specifies the optimization goal for $C_g$.
In SENSE, users set an upper bound $C_{g_{max}}$ for $C_g$, which the system tries to maintain for all tuples.
Each tuple has its individual coherence guarantee $C_g$ which may vary among tuples due to changing network conditions, processing delays on sensor nodes, or failures.
All these effects are reflected in the roundtrip time $\Delta$.

Assuming, that the transmission times between nodes follow an iid but otherwise arbitrary distribution, we expect the roundtrip time (the sum of values drawn from that random variable) to be approximately normally distributed.
In order to set $D_{max}$ such that a configurable fraction of tuples has $C_g \leq C_{g_{max}}$, we monitor point estimates for the mean $\bar\Delta$ and the standard deviation $\sigma$ of $\Delta$%
. The coherence guarantee follows a shifted normal distribution 
$C_g \sim \mathcal N(~\cdot~;\bar \Delta + \textrm{shift} , \sigma)$
as well, because the discussed sampling strategies are statistically unaffected by delayed tuples.
We set $D_{max}=C_{g_{max}}\!-\!3\sigma$ in order to gather tuples with
$C_g \leq C_{g_{max}}$ in 99.85\% of the requests according to the 68–95–99.7 rule~\cite{pukelsheim1994three}.  Monitoring $\bar \Delta$ and $\sigma$ has the negligible overhead of storing three floating point values.

We show the overall algorithm which tunes $\alpha$ and $\mu$, calculates $C_e$ and $C_g$, and emits result tuples in Algorithm~\ref{alg:onSampleArrivesAtLoopnode}.
This algorithm processes each sensor data tuple which returns to the loop node after passing the sensing loop.

\begin{algorithm}[t]
  \caption{Optimization of $\alpha$ and $\mu$ at the loopnode.}\label{alg:onSampleArrivesAtLoopnode}
\renewcommand{\algorithmicrequire}{\textbf{State:}}
\renewcommand{\algorithmicensure}{\textbf{Parameters:}}
\renewcommand{\Comment}[1]{{\color{blue}\scriptsize $\triangleleft$ #1}}
\begin{algorithmic}[1]
\Require
$l_s$,$N$,$t_l$,$s$,$\mu$,$p$
\Ensure
Tuple: $\langle t, \alpha_{min}, \alpha_{max}, t_{min}, t_{max}, v_1, \dots, v_N \rangle$
  \State $\Delta \gets l_s - $time()  \hspace{2.175cm}\Comment{compute roundtrip time}\label{line:computeRoundtripTime}
  \State $C_g \gets \Delta + \alpha_{max} - \alpha_{min}$ \hspace{1.005cm}\Comment{compute coherence guarantee}\label{line:computeCg}
  \State $C_e \gets t_{max} - t_{min}$  \hspace{1.7700cm}\Comment{compute coherence estimate}\label{line:computeCe}
  \State{emit($t, C_g, C_e, v_1, \dots, v_N$)}\hspace{8.900mm}\Comment{emit result tuple}\label{line:emitTuple}
  \If{$l_s \geq t_l$}\hspace{2.2920cm}\Comment{did earlier updates take affect?}\label{line:affectCheck}
    \State $\alpha \gets \Delta/2$ \hspace{24.1mm}\Comment{compute $\alpha$ without shift} \label{line:alpha}
      \State{$s \gets s + (p == \sign(D_{max}-C_g) ~?~ 1 : -1)$} \hspace{1mm}\Comment{exponent}\label{line:exponent}
      \State{$w \gets 2^s$}  \hspace{4.805cm}\Comment{step width}\label{line:stepWidth}
      \State{$\mu \gets \mu~/ \left(2 \sign\left(D_{max} - C_g\right) w \mu + 1 \right)$} \hspace{7.83mm}\Comment{new $\mu$}\label{line:mu}
      \If {$p==\sign(D_{max}-C_g)$}
      \State $p \gets  \sign(D_{max}-C_g)$ \hspace{0.65mm}\Comment{remember direction}
      \Else
        \State $p \gets 0$ \hspace{23mm}\Comment{half step size if direction changed}
      \EndIf
  \EndIf
\end{algorithmic}
\end{algorithm}

First, we compute the round trip time $\Delta$ for the tuple we process in Line~\ref{line:computeRoundtripTime}.
In Lines~\ref{line:computeCg} and~\ref{line:computeCe}, we compute $C_g$ and $C_e$ as discussed before.
We then emit the result tuple, including $C_g$ and $C_e$ in Line~\ref{line:emitTuple}.
The remainder of the algorithm updates $\alpha$ and $\mu$.
We update $\alpha$ and $\mu$ as soon as we observe the effect of previous updates.
In Line~\ref{line:affectCheck}, we check if the last $\alpha$ and $\mu$ update took affect for the tuple we process.
For example, if we request a sensor data tuple every 0.2s and the roundtrip time is 1s, this condition will hold from the fifth tuple after an update onwards.

\begin{figure}[t]
\centering
\includegraphics[width=\linewidth]{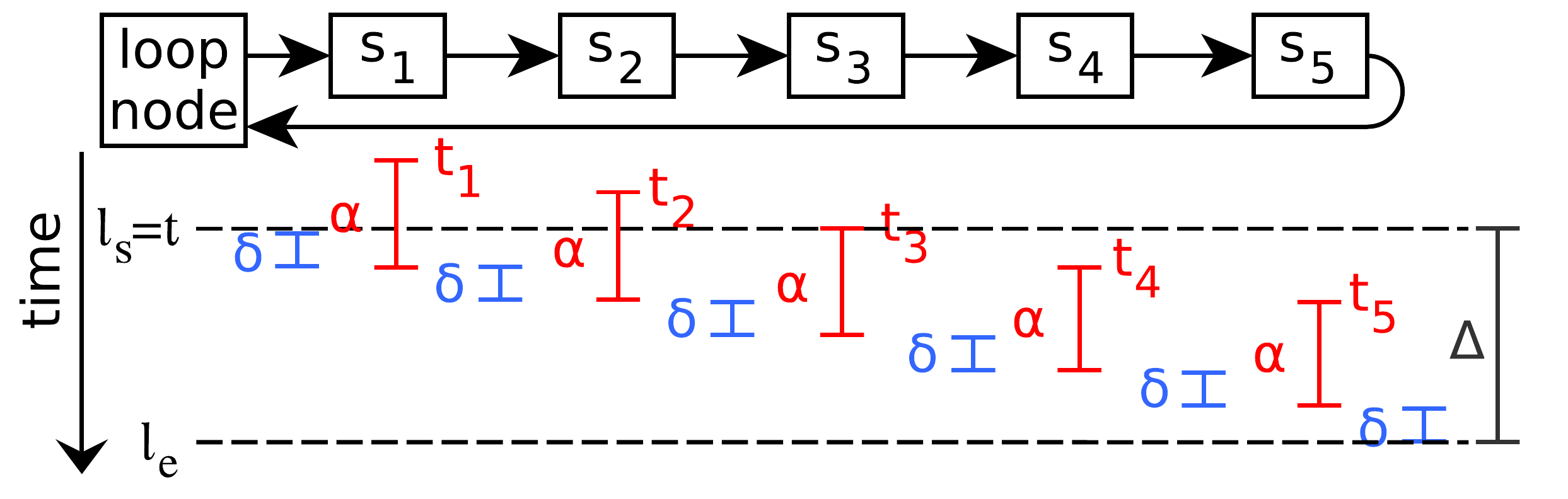}
\vspace{-6mm}
\caption{Selecting the optimal $\boldsymbol \alpha$=$\boldsymbol 3$ for $\boldsymbol \delta$=$\boldsymbol 1$ and $\boldsymbol N$=$\boldsymbol 5$.}
\label{fig:optalpha}
\end{figure}

\paragraph{Selecting $\boldsymbol{\alpha}$}
The optimum for the coherence guarantee is the roundtrip time $\Delta$ of a tuple.
This optimum ($C_g\!=\!\Delta$) is achieved if all sensor nodes provide values with an equal age $\alpha$ according to their local clocks at the join time $t_{now}$.
In general, an arbitrary age $\alpha$ leads to an optimal $C_g$ as long as the age is the same on all sensor nodes.
However, different $\alpha$ values imply smaller or larger read time deviations $\Delta_t$.
An optimal $\alpha$ entails the minimum mean squared deviation between $t$ and $t_{1},\dots,t_{N}$ (optimal $\Delta_t$). Our algorithm calculates the optimal $\alpha$ in Line~\ref{line:alpha} using Formula~\ref{eq:alpha}.
We derive Formula~\ref{eq:alpha} mathematically in Appendix~\ref{sec:derivationOfAlpha} and provide an example calculation in the following paragraph.
\begin{align}
  \alpha  = \frac{\delta (N+1)}{2}+(l_s-t) = \frac{\Delta}{2}+(l_s-t)\label{eq:alpha}
\end{align}
In Figure~\ref{fig:optalpha}, we show an example with $\delta$=$1$ and $N$=$5$, which leads $\alpha$=$3$.
We observe that half of the nodes select values read before $t$ and the other half selects values read after $t$.
This minimizes the mean squared error (i.e., difference) between $t$ and $t_1,\dots,t_N$.
Note that we set the request time $t$ equal to the loop start time $l_s$ in our example.
Differences between $t$ and $l_s$ would be added to $\alpha$ (rear part of Formula~\ref{eq:alpha}) which leads to the same mean squared error.
We add shifts to $\alpha$ whenever we send out a tuple from the loop node.

\begin{figure}[t]
\centering
  \includegraphics[width=0.7\linewidth]{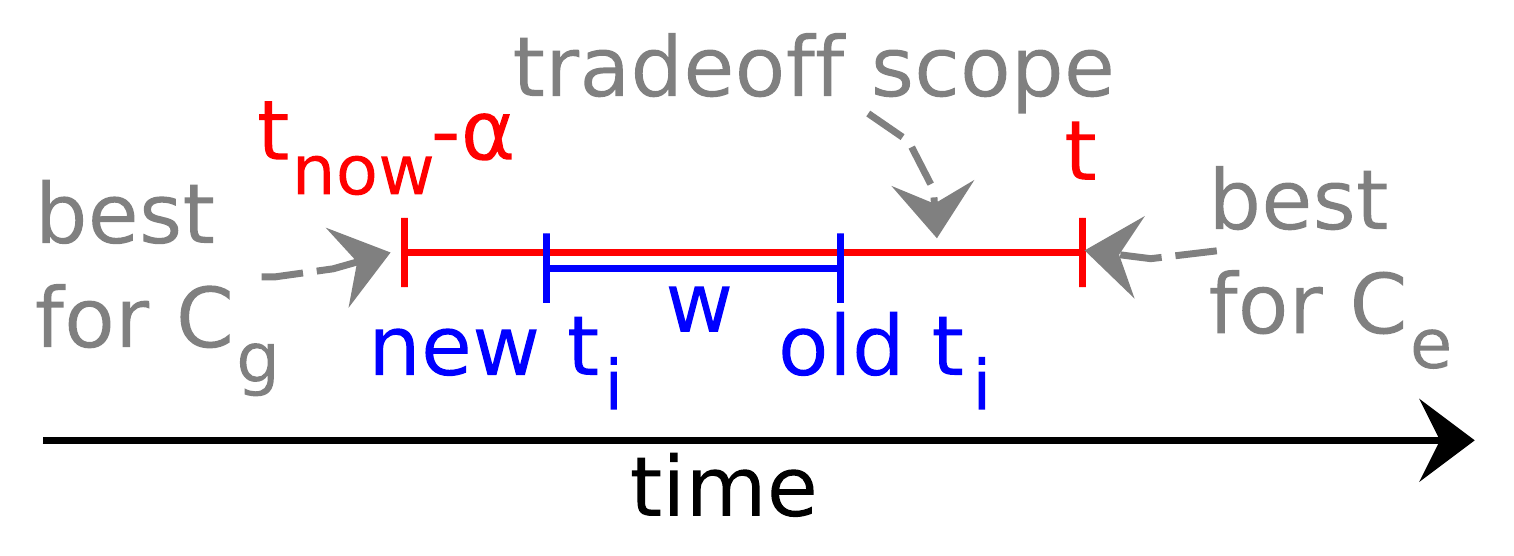}
  \caption{The step width $w$ in the tradeoff scope.}\label{fig:optrange}
\end{figure}

\paragraph{Selecting $\boldsymbol{\mu}$}
As discussed in Section~\ref{sec:sensor-node-algo}, sensor nodes select a value read between $t_{now}$-$\alpha$ (optimal $C_g$) and $t$ (optimal $C_e$) depending on $\mu$.
Figure~\ref{fig:optrange} illustrates this tradeoff scope.
Before we select $\mu$, we calculate the update step with $w$ with respect to the tradeoff scope.
Ideally, our result tuples have the best possible $C_e$ while not exceeding the desired upper bound for $C_g$ which we computed as $D_{max}$ (see above).

\textit{Update Step Width:}
First, we compute the exponent $s$ of the step with in Line~\ref{line:exponent} of Algorithm~\ref{alg:onSampleArrivesAtLoopnode}.
We then compute the step width $w$=$2^s$ in Line~\ref{line:stepWidth}.
In order to set the exponent $s$, we introduce an additional state variable $p$ which can have three stages depending on the previous run of the algorithm:
$1$ if $C_g>D_{max}$,
$-1$ if $C_g<D_{max}$, or
$0$ if $C_g=D_{max}$. Moreover, we set $p=0$ if $C_g>D_{max}$ changes to $C_g<D_{max}$ or vice versa.
We use $p$ to control when we decrement or increment the exponent $s$ of the step width $w$ (double or half $w$).

Figure~\ref{fig:covergingToDmax} shows an example where $C_g$ converges to $D_{max}$. For each step, we show $w$ and $p$ next to the curve separated by semicolon.
As long as $C_g$ approaches $D_{max}$, we increment $s$ (double $w$) in each step (up to Step 4).
This ensures that we reach $D_{max}$ fast.
As soon as we jump over the $D_{max}$ border, we change the direction, decrement $s$ (i.e., half $w$), and set $p=0$ (Step 5).
Setting $p=0$ ensures that the we decrement $s$ again for the next step (Step 6).
We always decrement $s$ twice after changing the direction.
This ensures convergence towards $D_{max}$.
We reach $C_g=D_{max}$ after 13 steps.

\begin{figure}[t]
\centering
  \includegraphics[width=\linewidth]{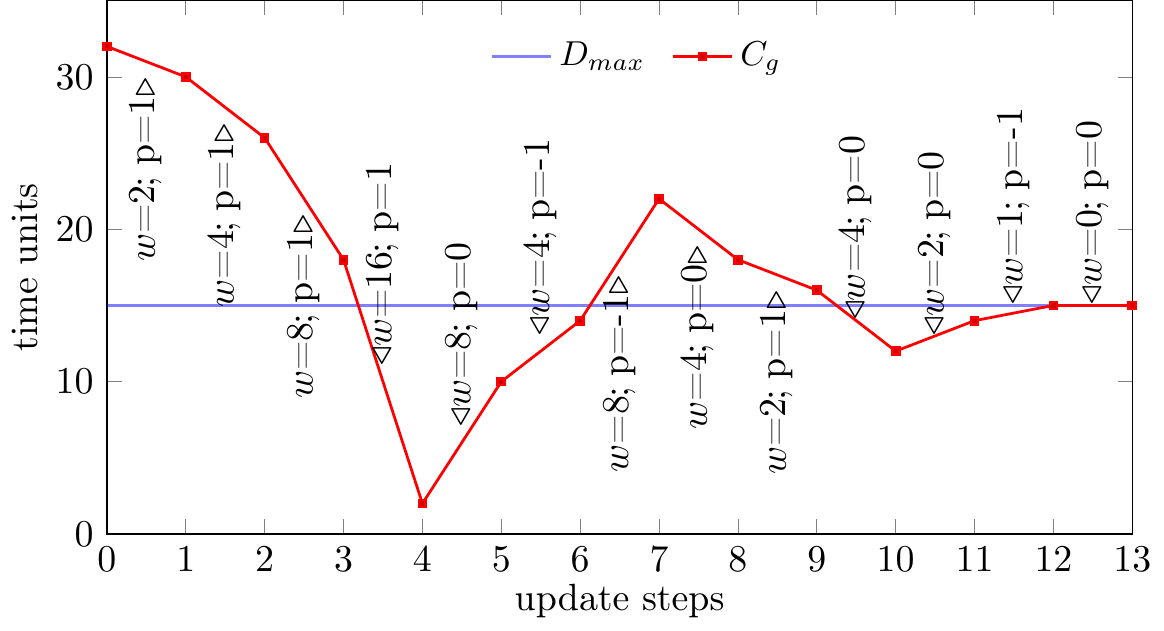}
  \captionof{figure}{Convergence of $\boldsymbol C_g$ to $\boldsymbol D_{max}$. (label: $\boldsymbol \triangleleft w ;\! p$)}\label{fig:covergingToDmax}
\label{fig:covergingToDmax}
\end{figure}

\textit{Update $\mu$:}
We derive a formula for $\mu$ such that changing $w$ changes the analytic optimum for $t_i$ by $w$ on sensor nodes (blue bar in Figure~\ref{fig:optrange}).
We update $\mu$ according to the desired step width $w$ with Formula~\ref{eq:mu} (Line~\ref{line:mu} in Algorithm~\ref{alg:onSampleArrivesAtLoopnode}).
\begin{align}
  \mu \gets \frac{\mu}{2 \sign(D_{max}-C_g) w \mu + 1}\label{eq:mu}
\end{align}
We derive this formula analytically in Appendix~\ref{sec:derivationOfMu}.
We transfer the parameter $\mu$ to sensor nodes instead of sending $w$ directly because we gain important flexibility.
By computing the optimal read time on the sensor node based on $\mu$, we can consider additional information in the optimization such as the timestamps available in the buffer and selected values of previous nodes.
This information is not available before the loop starts.
We explain these additional optimizations in Appendix~\ref{sec:opt-at-node}.
Moreover, we can weight deviations in $w$ flexibly on sensor nodes.
For example, we penalize deviations in the direction of $C_e$ with a higher order function than deviations in the direction of $C_g$.

\paragraph{Initialization of $\boldsymbol\alpha$ and $\boldsymbol\mu$}
We first acquire one tuple from sensors nodes with ad-hoc reads and, thereby, obtain $\Delta$, which we use to initialize $\alpha$ (Formula~\ref{eq:alpha}).
To initialize $\mu$, we estimate $t_{now}$ at any node $i$ based on $\delta$ (see calculation of $\alpha$).
We then set the first value of $\mu$ such that the coherence guarantee is just met in case $\delta$ does not change.
We derive the respective initialization formula in Appendix~\ref{sec:InitializationOfMu}.

\begin{figure}[t]
	\includegraphics[width=\linewidth]{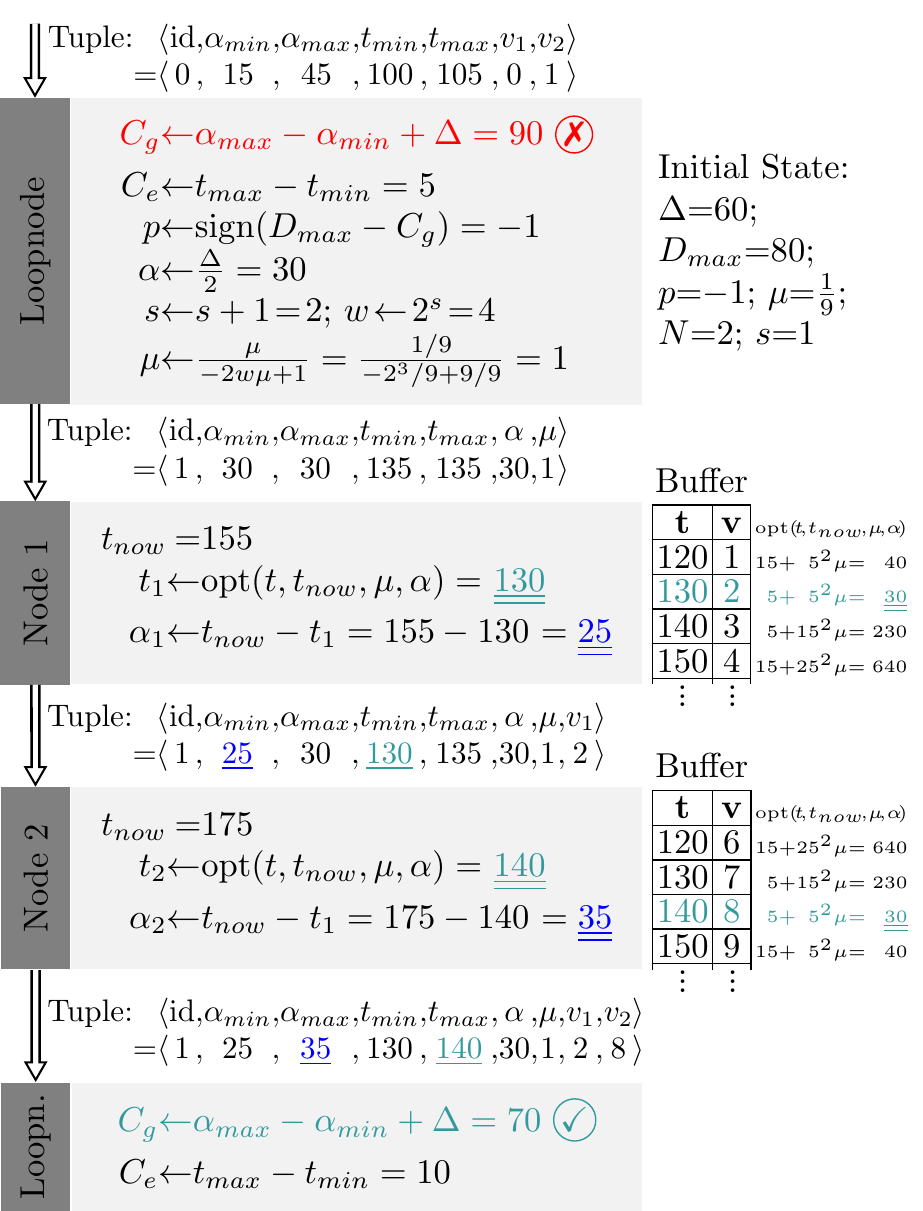}
	\caption{Example calculation: One optimization iteration for $\boldsymbol \alpha$ and $\boldsymbol \mu$ in a sensing loop with two nodes.}\label{fig:example-calculation}
\end{figure}%

\subsection{Example Calculation}%
We provide an example calculation in Figure~\ref{fig:example-calculation} which demonstrates how we tune $\alpha$ and $\mu$.
In this example, we optimize $\alpha$ and $\mu$ in a loop with two sensor nodes.

The loop node receives a result tuple (id 0) and computes the coherence guarantee of the tuple (red line).
We observe that $C_g > D_{max}$.
The loop node now computes $p$, $\alpha$, $s$, $w$, and $\mu$ with the formulas presented in the previous section.
At $t=135$, the loop node sends a request (id 1) to Node 1, which selects a value from its buffer according to Formula~\ref{eq:optFunc} presented in Section~\ref{sec:sensor-node-algo}.
Then, Node 1 forwards the tuple to Node 2 with updated $\alpha_{min}$ and  $t_{min}$.
Analogue to Node 1, Node 2 selects a value from its buffer and adjusts $\alpha_{max}$ and $t_{max}$.
Node 2 returns the result tuple to the loop node where we compute $C_g$ and $C_e$.
We observe that $C_e$ increased and $C_g$ reduces compared to the previous tuple.
We now achieved $C_g \leq D_{max}$ as intended.

\subsection{Splitting and Merging Sensing Loops}\label{sec:split-and-merge-loops}

If we either exceed $D_{max}$ or do not satisfy our latency requirements, the loopnode splits the set of sensor nodes into multiple sensing loops.
The maximum number of sensing loops is specified according to the loop node performance.

We split sensing loops such that that the roundtrip times of loops equal each other.
This isolates strugglers and therefore increases the global performance of the system.  The pipelines with the smallest roundtrip time are considered for a merge whenever the combined roundtrip time is expected to fulfill both the latency and coherence requirement.

\section{Scheduling Sensor Reads}\label{sec:scheduling-techniques}
SENSE uses three alternative techniques for scheduling sensor reads on sensor nodes to fill history buffers.

\para{Ad-hoc}
Reading ad-hoc is the simplest solution for scheduling sensor reads.
Whenever a tuple arrives at the sensor node, we read a value from the sensor and add that value to the tuple.
Ad-hoc reading requires no read scheduler and no buffer to store sensor values.
It solely optimizes for coherence guarantee $C_g$ because the age $\alpha$ is 0 for all values.
In exchange, the coherence estimate $C_e$ is high. 

\para{Periodic}
Scheduling sensor reads periodically (e.g., every 20ms) is the most common approach.
We gather values from sensors at a fixed frequency and store them in the history buffer.
Periodic scheduling is supported by almost all sensors and allows for optimizing $C_e$ and $C_g$ when selecting values from the buffer.
However, periodic scheduling may reads sensor values, which are never joined with any tuple.

\para{Schedule Next Read}
SENSE pre-schedules the next sensor read if possible.
Given the time of the next request, a sensor node can read a value exactly at the optimal time and store only that value in its history buffer.
Many algorithms which request sensor reads can provide their next request time up front which enables pre-scheduling. Examples are adaptive sampling techniques such as Adam~\cite{trihinas2015adam}, FAST~\cite{fan2012real}, and L-SIP~\cite{gaura2013edge} as well as on-demand scheduling techniques~\cite{traub2017ondemand}.
\textit{Schedule Next Read} reads required values only and, thereby, reduces the requires buffer size.
Since we can schedule sensor reads precisely, we can achieve the best results for $C_e$.

In order to compensate variations in the transmission time between sensor nodes in the optimization (Formula~\ref{eq:optFunc}), one can configure to populate the buffer with additional sensor values. Those values are sampled according to a configurable symmetric probability distribution centered around the scheduled read time.

\section{Failure Handling}\label{sec:failure-handling}

We now discuss fault-tolerance mechanisms for link outages, node outages, and buffer overflows.
We first introduce fallback nodes and then explain buffer overflow handling.

\subsection{Introducing Fallback Nodes}\label{sec:failure-handling:fbn}

We introduce fallback nodes to address link outages, node outages, and buffer overflows.
A fallback node replaces sensor nodes in case of failures and may be hosted redundantly on several servers, sensor nodes, or loop nodes.

If a sensor node cannot reach its succeeding node in the loop (missing TCP acknowledgement), it sends output tuples to the fallback node instead.
The fallback node then tries to reach the next available node in the loop and forwards the tuple to that available node skipping node(s) which are unavailable at the moment.
We remember unreachable nodes at the loop node and check periodically if they are back online.
These checks are performed asynchronously and do not delay processing tuples.
At first the fallback node tries to compensate for missing values with a cached value from the unreachable node.
If this fails, it tries to compensate for missing values with measures from an alternative sensor nearby.
If no alternative sensor or cached value is available the fallback node adds a null value.

It is desirable to collocate loop nodes and fall back nodes on the same server, because if collocated, a node outage can be seen as an \textit{implicit split} of the sensing loop. 
Since the tuple is returned to the loop node (=fallback node), the implicit split excerpts the same positive effect on the $C_g$-$C_e$-tradeoff as an intended loop split at the failing node and improves $C_g$ and $C_e$ accordingly as discussed in Section~\ref{sec:split-and-merge-loops}.
The loop node registers the deviating start time $t_s$ and incorporates it into the desired $\alpha$ value transmitted to the affected nodes.

In all cases, the loop node receives a final result tuple which it can forward directly. 
This is an advantage compared to central join topologies which cannot know if values are lost, late, or if the respective sensors are down.

\vspace{2mm}
\subsection{Managing Buffer Overflows}\label{sec:bufferManager}
\vspace{2mm}

Buffer Overflows on sensor nodes are another issue which we address with fallback nodes.
For example, a history buffer can overflow if a node does not receive requests for an unusual long time and, thus, cannot prune sensor values.

If a sensor node cannot store additional values in its buffer but expects requests which require additional values, it sends the oldest values to the fall back node and overwrites them with new values.
If a sensor node cannot reach any fallback node but needs to overwrite buffered values, follow up tuples which require overwritten data will fall back to the oldest value in the history buffer.
When a sensor node receives a tuple which requires values sent to the fallback node, it forwards the tuple to the fallback node which will add the respective values to the tuple and then forward the tuple to succeeding nodes in the loop.

In Algorithm~\ref{alg:bufferOverflow}, we show how we manage buffered values on sensor nodes.
Our buffer is a ring buffer, \texttt{pos} is the next write position, and $W_{join}$ and $W_{send}$ are two watermarks which keep track of processed tuples and values sent to the fallback node.
When we receive a tuple, we usually join it with a value from the sensor node buffer in the \textproc{JoinTuple} function.
The watermark $W_{join}$ is a timestamp which indicates which values have been sent to the fallback node.
If the request time $t$ of a tuple is smaller than $W_{join}$, the required sensor value has been sent to the fallback node (Line~\ref{line:sentToFallback}).
Otherwise, the required value is available in the sensor node buffer (Line~\ref{line:getFromBuffer}).
We finally set $W_{send}$ to remember that we processed all requests up to the request time $t$ of the processed tuple (Line~\ref{line:setWsend}).
We calculate $W_{send}$ such that we can overwrite values in the buffer which have timestamps larger than $W_{send}$ without sending them to the fallback node.

\begin{algorithm}[t]
  \caption{Buffer management on sensor nodes.}
  \label{alg:bufferOverflow}
\renewcommand{\algorithmicrequire}{\textbf{State:}}
\renewcommand{\algorithmicensure}{\textbf{Parameters:}}
\renewcommand{\Comment}[1]{{\color{blue}\scriptsize $\triangleleft$ #1}}
\begin{algorithmic}[1]
\Require
$\text{pos}$,$W_{join}$,$W_{send}$,$\text{buffer}$
  \Function{JoinTuple}{inputTuple}
    \If {$\text{inputTuple.t}<W_{join}$}
  	  \State Forward input tuple to fallback node.\label{line:sentToFallback}
    \Else
      \State Select best $v_i$ from buffer with Formula~\ref{eq:optFunc}.\label{line:getFromBuffer}
      \State Send result tuple to the next node.
    \EndIf
    \State $W_{send} \gets \min(t_{now}-\alpha,t)$\label{line:setWsend}
  \EndFunction
  \Function{ReadValue}{~\!}
    \If {$\text{buffer[pos].t}>W_{send}$}\label{line:checkWsend}
  	  \State Send buffer[pos] to fallback node.
  	  \State $W_{join} \gets (\text{buffer[pos].t}+\text{buffer[pos-1].t}) / 2$\label{line:setWjoin}
    \EndIf
    \State $\text{buffer[pos]} \gets \text{sensor.read()}$
    \State $\text{pos} \gets \text{pos}+1$
  \EndFunction
\end{algorithmic}
\end{algorithm}

In the function \textproc{ReadValue}, we add a new sensor value to the buffer which overwrites an old value in the buffer.
In Line~\ref{line:checkWsend}, we check if we need to send the value we overwrite to the fallback node.
This is the case, if we expect requests which potentially require the value we overwrite.
If we send a value to the fallback node, we set $W_{join}$ accordingly in Line~\ref{line:setWjoin}.
$W_{join}$ specifies which requests are sent to the fallback node and which requests are processed locally.
If the optimal read time for a request is closer to the value we sent to the fallback node, we also send the request to the fallback node.
Otherwise, we process the request locally.

\section{Evaluation}\label{sec:evaluation}
We first present our experiment setup. Then, we demonstrate SENSE on our testbed, before we evaluate time coherence optimization, followed by an analysis of throughput, latency, and CPU utilization. Finally, we evaluate SENSE for a large scale parameter space to show its general applicability. %

\subsection{Experiment Setup}
We implement the Loop Node and Sensor Node components of SENSE in C++ libraries which can easily be integrated in different host applications and systems.
We transmit 64bit timestamps with nanosecond precision and 64bit values as payload.
Note that the type of acquired sensor data does not affect the outcome of our experiments because our algorithms do not process the acquired sensor values.

\textbf{Sensor Node Testbed.} Our testbed consists of 10 Raspberry Pi 3B+, which are connected to a TP-Link TL-SG116 16-Port Switch, which is connected to a Netgear R6120 Router.
We collocate a loop node and a fall back node on one Raspberry Pi as discussed in Section~\ref{sec:failure-handling:fbn}. On each Raspberry Pi, we run 10 sensor node instances (100 logical sensor nodes in total). We ensure that two succeeding nodes in a loop are never located on the same physical node. All messages are sent from their source node via the switch to the router and back via the switch to the receiver (regular TCP/IP routing).

\textbf{Large Scale Experiments.} To enable a long term evaluation of SENSE, with large numbers of sensors and many different configurations, we also run SENSE in a simulation.
Our simulation uses the same SENSE libraries as our testbed, but allows for fast forward execution.
We use the the NS-3 network simulator in our experiments to emulate realistic network delays, jitter, and transmission failures~\cite{riley2010ns}.

\textbf{Performance Measures.}
We evaluate throughput, latency, and CPU-load on one core of an Intel Core i7-7600U CPU with 2.80GHz and 4MB cache on a computer with 15GB main memory.
We monitor the CPU usage with \textit{getrusage}~\cite{getrusage} and show sustained throughput~\cite{karimov2018benchmarking}.
We use a multi-core system to not affect performance measures by monitoring software running on the same system.

\subsection{SENSE System Demonstration}\label{sec:sense-system-demo}
\begin{figure}[t]
\centering
\includegraphics[]{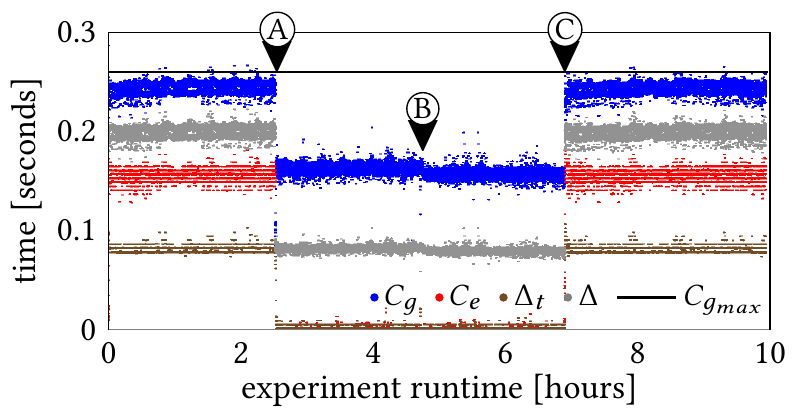}
\caption{Impact of node failures on the coherence tradeoff with collocated Loop- and Fallback nodes.}
\label{fig:fault-tol}
\end{figure}

In this section, we demonstrate SENSE on our sensor node testbed in a 10 hour experiment and evaluate the fault tolerance of the system.
We show the results of the experiment in Figure~\ref{fig:fault-tol}. The plot shows the coherence guarantee ($C_g$), the coherence estimate ($C_e$), the read time deviation ($\Delta_t$), and the roundtrip time ($\Delta$) for each tuple which passes the sensing loop. In addition, we show the desired upper limit $C_{g_{max}}$ for $C_g$.
Recall that the user-defined incoherence limit $C_{g_{max}}$ is transformed to a system internal threshold $D_{max}$, which adapts to network conditions (Section~\ref{sec:loop-node-algo}).

We analyze three events in Figure~\ref{fig:fault-tol}: At time \circled{\small A}, the 48th sensor in the loop crashes,
at time \circled{\small B}, the 49th and 50th node crash, and
at time \circled{\small C}, all crashed nodes recover.

Initially, before \circled{\small A}, we see that the system properly sets $D_{max}$ and adjusts the $C_e$-$C_g$-Tradeoff to minimize $C_e$ while maintaining $C_g<C_{g_{max}}$ for 99.5\% of all tuples.
When the 48th node in our loop consisting of 100 nodes crashes (time \circled{\small A}), the 47th node sends its tuples to the fallback node instead.
The fallback node forwards the tuples to the 49th node.
Since loop node and fallback node are collocated, we see an improvement of $C_g$ and $C_e$, as the failure causes an \textit{implicit split} as discussed in Section~\ref{sec:failure-handling:fbn}. In exchange, there is a slightly higher workload at the loop node, which -- in this scenario -- the node is able to handle without increasing the latency.
When the succeeding 49th and 50th node fail, we only see a minor improvement of $C_g$ and $C_e$.
Skipping those two nodes reduces the roundtrip time $\Delta$, but does not cause another \textit{implicit split} as the loop is already split at the 48th node.
At time \circled{\small C} all crashed nodes recover and the systems reintegrates them into a single sensing loop as it was before the first node outage.

\textbf{Discussion.}
We observe that SENSE is able to deal with node outages and to reintegrate recovering nodes. The system ran for ten hours before we terminated the experiment.

\subsection{Optimizing Time Coherence}\label{sec:exp-opt-coherence}
\begin{figure}[t]
\centering
\captionsetup[subfigure]{justification=centering,position=t}
\begin{subfigure}[b]{0.68\linewidth}%
  \hspace{-4mm}  
  \includegraphics[]{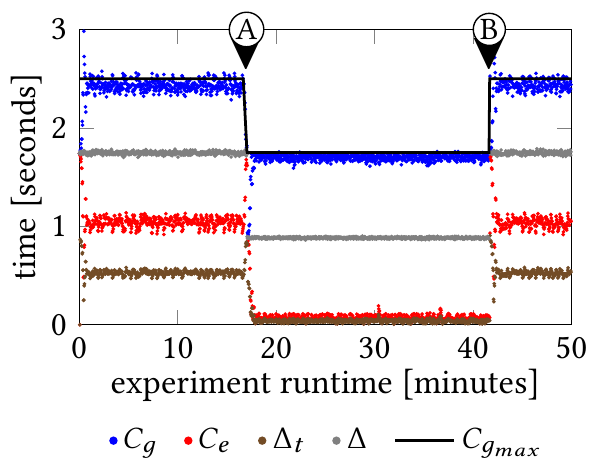}\vspace{0.5mm}
  \caption{Result with scheduled sensor reads.}
  \label{fig:triplePlot-SNR}
\end{subfigure}%
\begin{subfigure}[b]{0.25\linewidth}%
\includegraphics[]{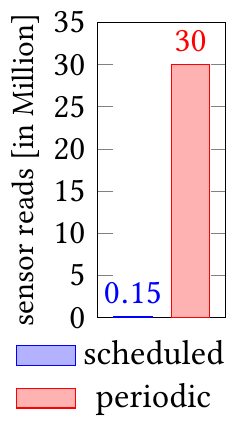}
\caption{Number of sensor reads.}
\label{fig:triplePlot-count}
\end{subfigure}%
\caption{Evolution of coherence guarantees and estimates for changing incoherence limits ($\boldsymbol D_{max}$).}
\label{fig:exp-triple-plot}
\end{figure}

In this experiment, we evaluate the time coherence optimization introduced in Section~\ref{sec:tolerances}.
We analyze two scenarios: In Scenario \circled{\small A}, we decrease $C_{g_{max}}$ (better guarantee, worse estimate). In Scenario \circled{\small B}, we increase $C_{g_{max}}$ (worse guarantee, better estimate).
In our experiment, we analyze the following aspects:
\begin{enumerate*}
	\item The adaptivity with respect to changing coherence requirements.
	\item The optimization of the $C_e$-$C_g$-tradeoff with respect to $D_{max}$.
	\item The functionality of loop splits and merges.
	\item Different approaches for scheduling sensor reads.
\end{enumerate*}

\paragraph{Setup}
We show the results for data transmissions via LTE~\cite{piro2011lte, riley2010ns}.
We also tested Wifi connections which had similar but less volatile transmission times.
LTE has 5-10 times higher transmission times than LAN (Section~\ref{sec:sense-system-demo}).
In this experiment, we simulate a sensing loop with 200 sensor nodes.

\paragraph{Scenario A: Decreasing $\boldsymbol D_{max}$}
In Figure~\ref{fig:triplePlot-SNR} at time \circled{\small A}, we change the system internal threshold $D_{max}$ such that $D_{max}$ is below the roundtrip time $\Delta$, which forces SENSE to adapt by splitting the sensing loop.

The gray dots in Figure~\ref{fig:triplePlot-SNR} show the roundtrip times $\Delta$ of tuples in the sensing loop.
As explained in Section~\ref{sec:CgandCe}, the minimum value for $C_g$ is $\Delta$.
Since we reduce $D_{max}$ slightly below $\Delta$ in our experiment, SENSE correctly detects that we cannot achieve $C_g \leq D_{max}$ with a single loop covering all 200 sensor nodes and splits the loop accordingly.
As a consequence, the coherence guarantee $C_g$ converges below the new value of $D_{max}$, which shows that SENSE correctly identified that it needs to adjust the sensing loop. At the same time, SENSE immediately reduces the coherence estimate $C_e$ as much as possible under the constraint $C_g \leq D_{max}$. %

\paragraph{Scenario B: Increasing $\boldsymbol D_{max}$}
In Figure~\ref{fig:triplePlot-SNR} at time \circled{B}, we change the system internal threshold $D_{max}$ back to its initial value, which enables SENSE to adapt by merging sensing loops.
SENSE correctly detects that we can now fulfill the constraint $C_g \leq D_{max}$ with a single loop covering all 200 sensor nodes and merges the two sensing loops accordingly.
As a consequence, the coherence guarantee $C_g$ converges below the new value of $D_{max}$.

\paragraph{Impact of Read Scheduling}
We executed the experiment with two read scheduling techniques, \emph{Schedule Next Read} (Figure~\ref{fig:triplePlot-SNR}) and \emph{Periodic Scheduling} (not shown).
The resulting plots are identical, except that \emph{Schedule Next Read} achieves smaller deviations from the desired read times ($\Delta_t$), because we can schedule sensor reads at optimal times with respect to future request times.
For reference, we include the plot for periodic sampling in Appendix~\ref{app:periodic_scheduling}.
In Figure~\ref{fig:triplePlot-count} we observe, that \emph{Schedule Next Read} reduces the number of required sensor reads by more than 99\% compared to periodic scheduling because we can schedule exactly one sensor read per sensor and requested tuple.

\paragraph{Discussion}
We observe that SENSE optimizes the coherence estimate of tuples while keeping the coherence guarantee within a user-defined upper limit.
Thereby, SENSE adapts quickly to changes (network conditions and coherence requirements) and splits and merges sensing loops as required.
Pre-scheduling sensor reads reduces the number of required reads drastically compared to reading values periodically.

\subsection{Throughput, Latency, and CPU Load}\label{sec:exp-perfomance}
\begin{figure}[t]
\captionsetup[subfigure]{justification=centering,position=t}
\begin{subfigure}[t]{0.33\linewidth}%
\centering
  \includegraphics[]{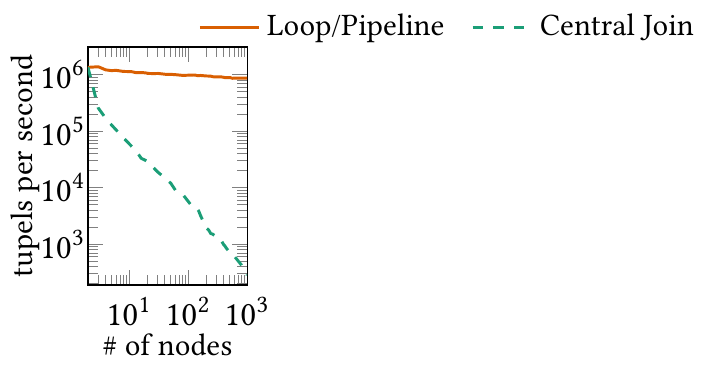}
  \caption{ Throughput.}\label{fig:throughput}
\end{subfigure}%
~~~~~~%
\begin{subfigure}[t]{0.33\linewidth}%
\centering
  \includegraphics[]{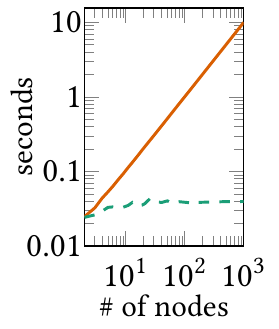}
  \caption{Latency.}\label{fig:latency}
\end{subfigure}%
\begin{subfigure}[t]{0.33\linewidth}%
	\centering
  \includegraphics[]{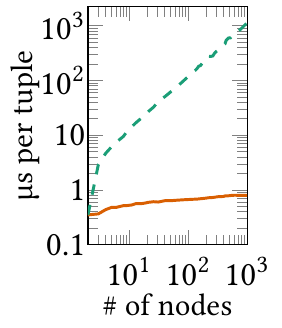}
  \caption{ CPU time.}\label{fig:cputimepertuple}
\end{subfigure}%
\caption{Performance Evaluation on the Loop Node.}
\label{fig:exp-performance}
\end{figure}

In order to evaluate the performance of a sensing loop, we measure throughput, latency, and CPU utilization of loop nodes, which are the bottleneck for these measures.
Individual sensor nodes in the same sensing loop always face the same load independent of the number of nodes in the loop. %
The latency includes the network latency for transmitting values from sensors to a central node (central join) and for transmitting values in a loop or pipeline.

\paragraph{Throughput}
We observe in Figure~\ref{fig:throughput} that a central join solution does not scale to thousands of sensors like we expect in upcoming Internet of Things applications.
The throughput decreases for larger numbers of sensors because matching values form many sensors to coherent tuples centrally requires many timestamp lookups and comparisons.
Our pipeline-based solution overcomes this problem because it drastically reduces the number of inputs which need to be joined at a central node.

\paragraph{Latency}
Figure~\ref{fig:latency} shows the latency between the desired read time of a tuple and the time the systems returns the tuple to the user.
The latency in sensing loops scales linearly with the number of sensor nodes (i.e. with the number of hops).
A central join topology has a much smaller latency than a sensing loop because it requires only one hop from sensors to the central node.
In general, transmission times dominate the latency of both approaches rather than computation times.
Please note that our experiment shows the latency with sustained throughputs.
Thus, central joins achieve smaller latencies with a much smaller throughput than sensing loops.

\paragraph{CPU Utilization}
Sensing loops reduce the CPU utilization at a central node (Figure~\ref{fig:cputimepertuple}) significantly compared to central join topologies.
Since the sensing loop provides a complete sensor data tuple, the remaining computation at the loop node is limited to calculating $C_e$, $C_g$, $\alpha$, and $\mu$ (remember Algorithm~\ref{alg:onSampleArrivesAtLoopnode}).
In contrast, a central join of individual sensor values must match each incoming value with values of other sensors and check if it can output a complete tuple.

\paragraph{Discussion}
Central join topologies are limited
by their CPU performance which limits the sustained throughput.
Sensing loops are limited by their latency which results from accumulated hop times between sensor nodes.
The strength of our solution is the adaptive combination of both approaches: multiple sensing loops combined by a central join.
We can see in Figure~\ref{fig:exp-performance} that a sensing loop can acquire values from 100 sensors with less than two seconds latency.
A central join can combine inputs from 100 sensing loops with a throughput of 5000 tuples per second.
Hence, we can acquire values from 10000 sensors with a latency below 3 seconds and guaranteed time coherence below 2 seconds.

\subsection{Coherences and Read Time Deviations}
\begin{figure}[t]
\captionsetup[subfigure]{justification=centering,position=t}
\begin{subfigure}[t]{0.33\linewidth}%
\centering
  \includegraphics[]{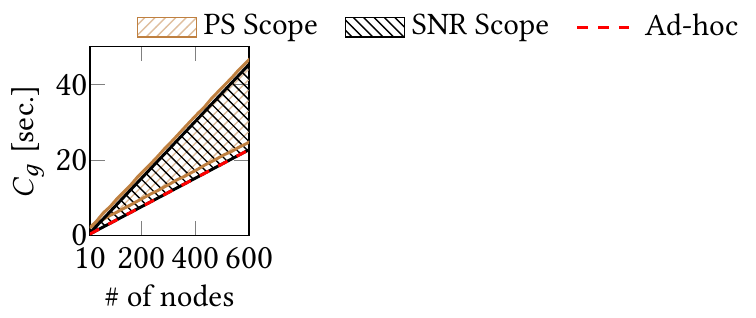}
  \caption{ Coherence Guarantee.}\label{fig:cg-scope}
\end{subfigure}%
~~~~~~%
\begin{subfigure}[t]{0.33\linewidth}%
\centering
  \includegraphics[]{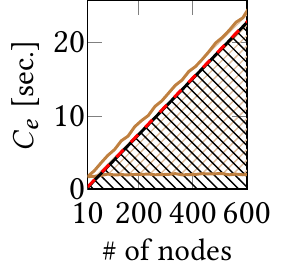}
  \caption{ Coherence Estimate.}\label{fig:ce-scope}
\end{subfigure}%
\begin{subfigure}[t]{0.33\linewidth}%
	\centering
  \includegraphics[]{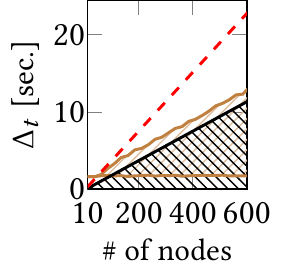}
  \caption{ Readtime Deviation.}\label{fig:dt-scope}
\end{subfigure}%
\caption{Evaluation of coherence measures.}
\label{fig:exp-coherence-measures}
\end{figure}

In this experiment, we analyze the impact of the number of nodes in one sensing loop on coherence guarantees, coherence estimates, read time deviations, and the related tradeoffs.
In Figure~\ref{fig:exp-coherence-measures}, we show the optimization scopes for coherence guarantees (Figure~\ref{fig:cg-scope}), coherence estimates (Figure~\ref{fig:ce-scope}), and read time deviations (Figure~\ref{fig:dt-scope}).
We simulate average hop times of 38ms between sensors nodes and select the loop start time $l_s$ as request time $t$ for tuples.
We show the optimization scopes for periodic scheduling~(PS), schedule next read~(SNR), and ad-hoc reading.

\begin{figure*}
	\begin{subfigure}[b]{0.33\linewidth}
		\includegraphics[
			width=\linewidth,
			trim=0 0 0 23mm,
			clip]
			{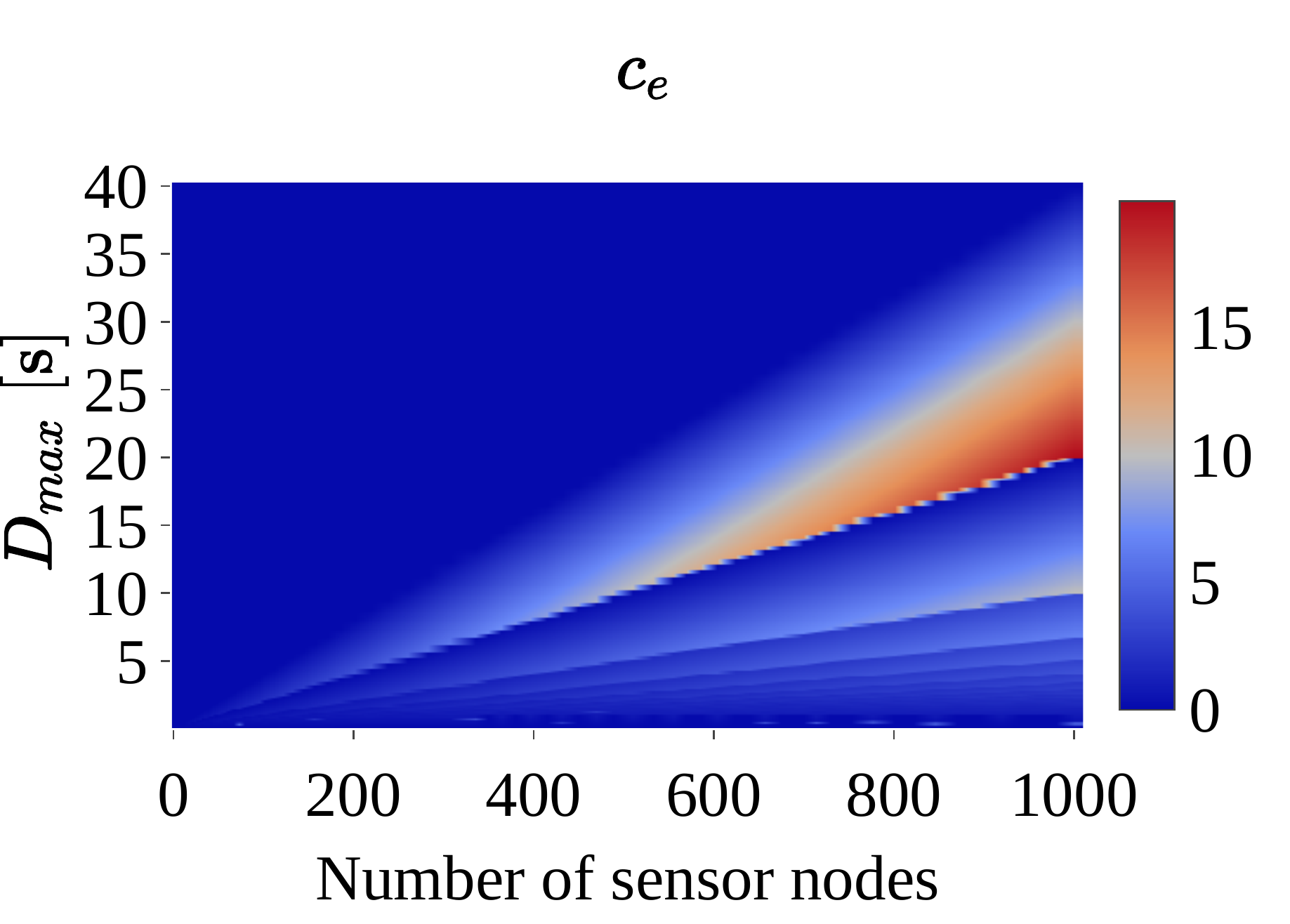}
		\caption{Coherence Estimate ($\boldsymbol C_e$)}
		\label{fig:heatmaps-ce}
	\end{subfigure}
	\begin{subfigure}[b]{0.33\linewidth}
		\includegraphics[
			width=\linewidth,
			trim=0 0 0 23mm,
			clip]
			{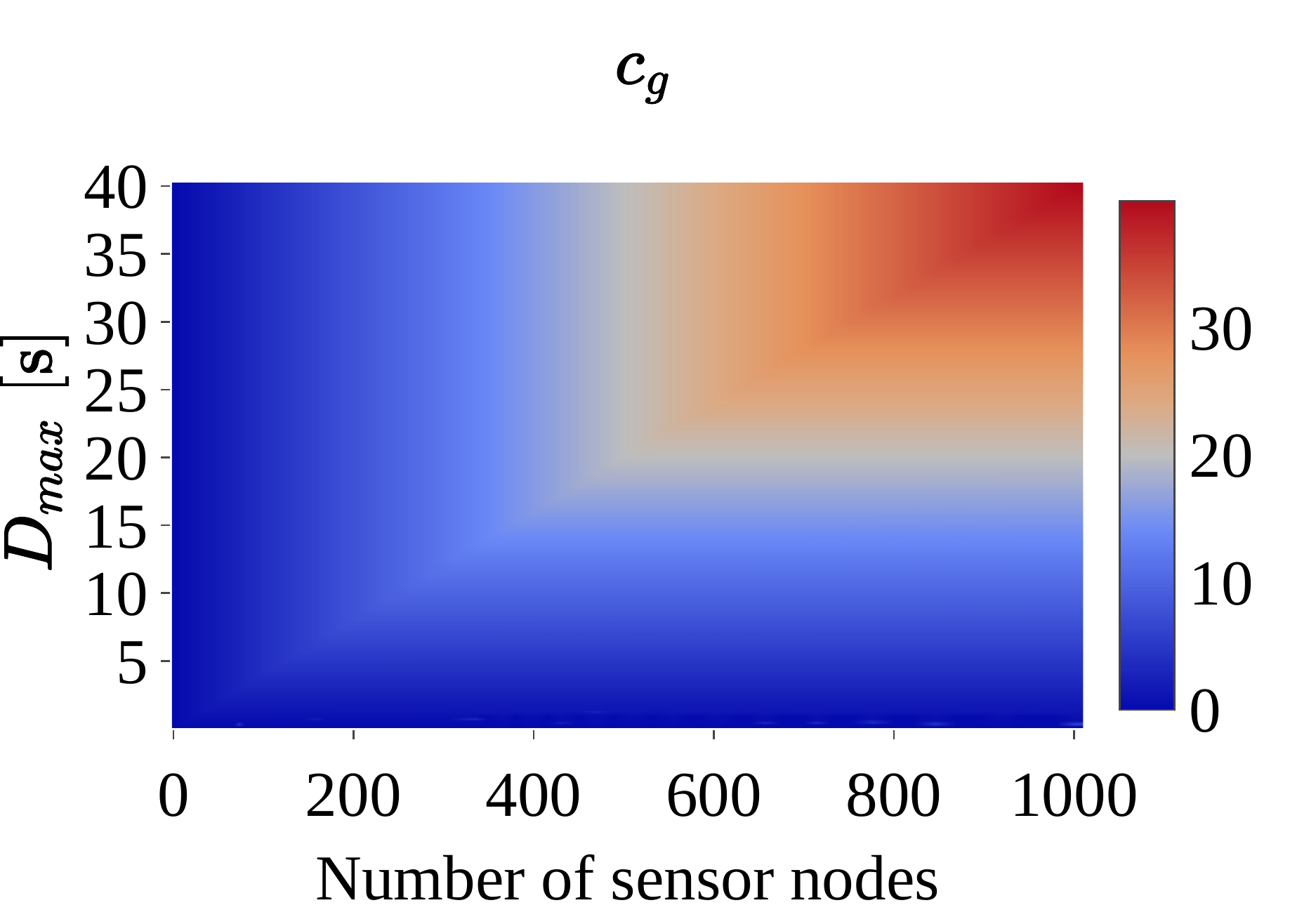}
		\caption{Coherence Guarantee ($\boldsymbol C_g$)}
		\label{fig:heatmaps-cg}
	\end{subfigure}
	\begin{subfigure}[b]{0.33\linewidth}
		\includegraphics[
			width=\linewidth,
			trim=0 0 0 23mm,
			clip]
			{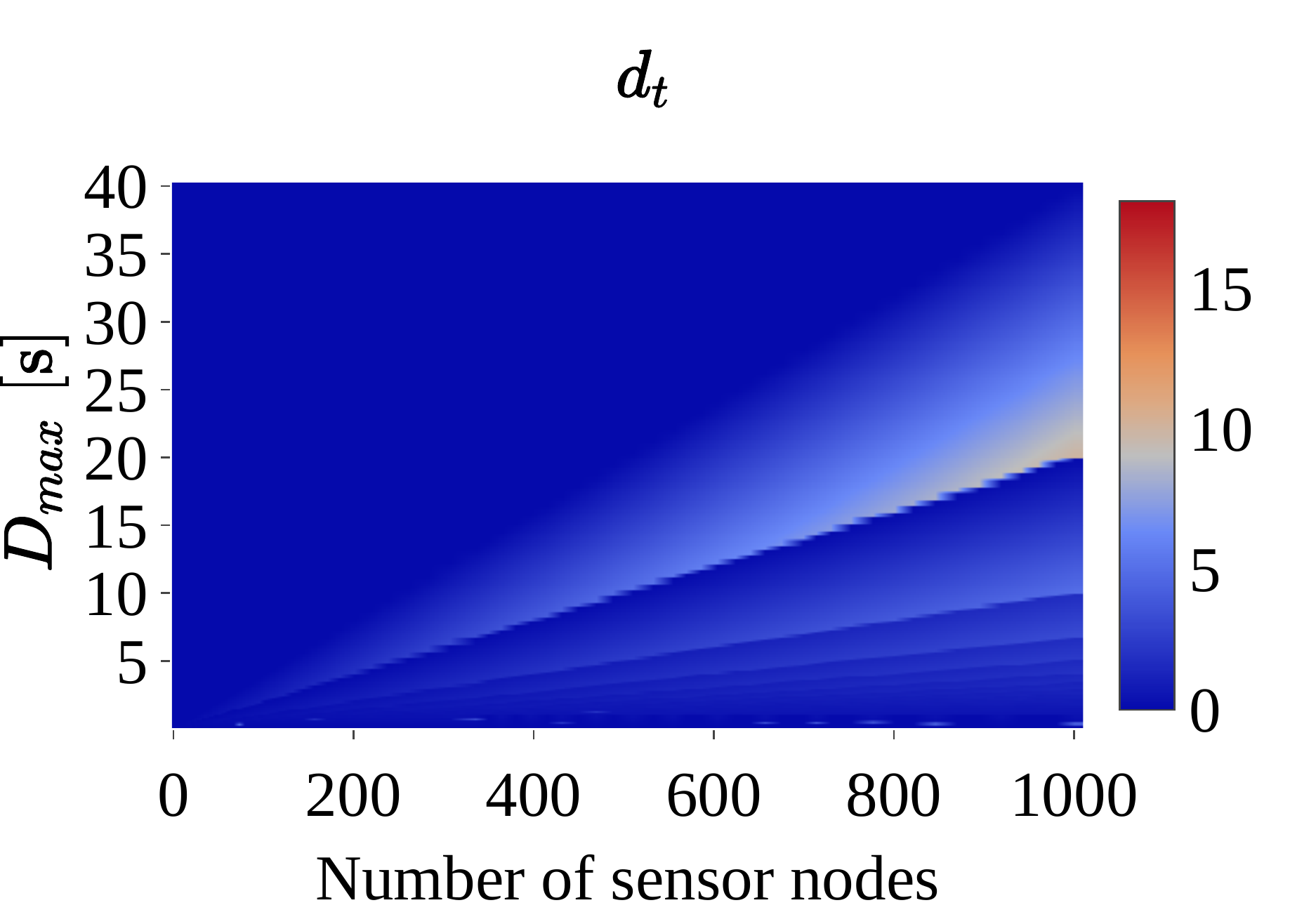}
		\caption{Read Time Deviation ($\boldsymbol \Delta_t$)}
		\label{fig:heatmaps-dt}
	\end{subfigure}
	\caption{Coherence measures depending on loop lengths and $\boldsymbol D_{max}$.}
	\label{fig:heatmaps}
\end{figure*}

\paragraph{Coherence Guarantee}
The smallest possible coherence guarantee is the roundtrip time $\Delta$ which increases linearly with the loop length.
We achieve this optimum if all nodes provide a value with the same age $\alpha$.
With SNR, we can achieve the optimal coherence guarantee because we can schedule sensor reads precisely with respect to following requests.
Ad-hoc reading always results in $\alpha$=$0$ on all nodes which also leads to an optimal coherence guarantee.
Periodic Scheduling (PS) leads to slightly worse guarantees, because it adds a deviation between the optimal read time and the read time which is available in the history buffer.
The worst case coherence guarantee results from solely optimizing for coherence estimates, which means that we select values read exactly at the request time $t$ (according to sensor node clocks).
In this case, the difference between $\alpha_{min}$ and $\alpha_{max}$ equals $\Delta$ and the coherence guarantee increases to $2\Delta$ accordingly.

\paragraph{Coherence Estimate}
The optimal coherence estimate results from selecting values which were read exactly at the request time $t$ as described above.
We can achieve this optimum with SNR because we can schedule sensor reads precisely at future request times.
Periodic reading adds a deviation between the optimal read time and the read time available in the sensor node buffer which explains the slight shift of the optimization scope.
The largest coherence estimate results from optimizing solely for the coherence guarantee.
In this case, $\alpha_{min}$=$\alpha_{max}$ and $C_e\!\!\approx\!\!\Delta$ assuming correctly synchronized sensor node clocks.
Ad-hoc reading results in the worst-case coherence estimate because it reads when it receives tuples without considering the request time.

\paragraph{Read Time Deviation}
The read time deviation $\Delta_t$ is the maximum difference between the request time $t$ and any read time of a value contained in the result tuple.
We regularly tune the parameter $\alpha$ to minimize $\Delta_t$ (remember Figure~\ref{fig:tradoff-ce-cg}).
Figure~\ref{fig:dt-scope} shows the impact of this optimization.
Both periodic scheduling and schedule next read achieve $\Delta_t\!\approx\!C_e/2$, which is the optimum.
In contrast, reading ad-hoc doubles $\Delta_t$ because of the missing $\alpha$-optimization.

\paragraph{Discussion}
Our experiment illustrates the best-cases and worst-cases for $C_e$, $C_g$, and $\Delta_t$ as well as the scope of the $C_e$-$C_g$-tradeoff.
Schedule next read (SNR) achieved slightly better results because of precise scheduling of sensor reads with respect to future requests.
A naive solution which reads ad-hoc achieves an optimal $C_g$, at the cost of a worst-case $C_e$ and a doubled read time deviation.
In contrast, our solution adapts the tradeoff between $C_e$ and $C_g$ flexibly with respect to user-defined coherence requirements.

\subsection{Large Scale Parameter Exploration}

In this experiment, we evaluate whether our results carry over to a large parameter space.
We provide a quantitative analysis of the impact of the number of sensor nodes in a sensing loop and the coherence requirements ($D_{max}$) on the coherence measures ($C_g$, $C_e$, and $\Delta_t$).
For each measure, we show a dedicated heatmap in Figure~\ref{fig:heatmaps}. %
Overall, the heatmaps show the results of 57600 experiments with a simulated duration of 66 minutes per experiment and a mean hop time of 20ms.
We show $D_{max}$ on the y-axis, the amount of sensors on the x-axis, and coherence measures as heatmap colors.

\paragraph{Coherence Estimate}
In Figure~\ref{fig:heatmaps-ce}, we show the coherence estimates depending on $D_{max}$ and the number of sensors.
We observe diagonal lines which correspond to loop splits.
Above these lines, the system optimizes strongly for $C_g$ to keep $C_g \leq D_{max}$.
In exchange, $C_e$ increases which explains the increasing values for $C_e$ when approaching the split.
Directly below the lines which indicate loop splits, the system can optimize strongly for $C_e$ without violating $C_g \leq D_{max}$ which results in the best (i.e., smallest) values for $C_e$.

\paragraph{Coherence Guarantee}
In Figure~\ref{fig:heatmaps-cg}, we show the coherence guarantees for the same experiments.
One can observe that pipeline splits are barely visible in this plot.
This shows that our system always utilizes $D_{max}$ to relax $C_g$ and to optimize for $C_e$.
At the same time, we observe that our system does not violate $D_{max}$.
For smaller numbers of sensors, the system does not need to fully utilize $D_{max}$ for achieving an optimal $C_e$, which explains the blue area in the upper left corner of Figures~\ref{fig:heatmaps-ce} and~\ref{fig:heatmaps-cg}.

\paragraph{Read Time Deviation}
Figure~\ref{fig:heatmaps-dt} shows the read time deviation $\Delta_t$ for our experiments.
Similar to the results presented in the previous section, we achieve $\Delta_t\!\approx\!C_e/2$, which is the optimal result for $\Delta_t$. This proves that the loop node chooses the optimal value for $\alpha$.

\paragraph{Discussion}
The presented heatmaps verify that SENSE achieves the desired results not only for selected combinations of parameters, but for a wide range of \mbox{setups}.
Our solution optimizes $C_e$ without violating \mbox{$C_g \leq D_{max}$}.
Pipeline splits take place as required to ensure this behavior (Figure~\ref{fig:heatmaps-ce}).
The system utilizes the upper limit for coherence guarantees ($D_{max}$) to relax $C_g$ and to optimize $C_e$ as much as possible (Figure~\ref{fig:heatmaps-cg}).
The read time deviation is about $C_e/2$, which proves an optimal selection of $\alpha$ (Figure~\ref{fig:heatmaps-dt}).

\section{Related Work}\label{sec:related-work}
In this section, we discuss related works not discussed before.

Time coherence was studied in different contexts before.
Srinivasan et al. discuss the \textit{temporal coherence} of virtual data warehouses which work as cache for data sources~\cite{srinivasan1998maintaining}.
Deolasee et al. propose an adaptive push-pull method to maintain the coherence of such cached data copies efficiently~\cite{deolasee2001adaptive}.
Agrawal et al. discuss techniques to smartly select up to date (temporal coherent) cached data copies for serving client requests~\cite{agrawal2004construction}.
These works consider the coherence between data sources and several copies of these sources.
In contrast, our work focuses on acquiring time coherent tuples which contain values from many distributed sensor nodes.

TiNA addresses temporal coherence-aware in-network aggregation~\cite{beaver2003power, sharaf2003tina}.
Sensors transmit new readings only if values changed more than a given tolerance.
This allows for trading off energy consumption in sensor networks against the quality of data~\cite{sharaf2004balancing}.
Similarly, adaptive sampling techniques~\cite{gaura2013edge,fan2012real,trihinas2015adam} monitor the change of sensor values and transmit new measurements only if they are sufficiently different from previous ones.
However, these techniques do not provides time coherence guarantees for result tuples. %

The term coherence was also used in the context of fusing data from several sensors (i.e., sources) to one logically coherent data base~\cite{kokar1993Review,luo1989Multisensor,luo2002Multisensor}.
This multi-sensor fusion is independent of time coherence and builds on the logical consistence of values from different sources instead.

Diverse works propose clock synchronization techniques or synchronization optimizations~\cite{kulkarni2014logical,lasassmeh2010time, leng2010on,Misra2017Enabling,ping2003delay,sichitiu2003simple,sivrikaya2004time} and/or measure round trip times of synchronization messages~\cite{cristian1989probabilistic, li2006global}.
In this paper, we use synchronized clocks for coherence estimates wherever possible.
However, in the IoT, it is impractical to ensure complete synchronization among all clocks.
To this end, we provide additional coherence guarantees to detect and handle out-of-sync sensor node clocks.

Madden et al. introduce \textit{acquisitional query processing} and show the benefits of interleaving data gathering operations with data manipulation operations in TinyDB~\cite{madden2005tinydb,madden2002tag}.
Shrivastava et al. extend the scope of TinyDB with complex aggregations and range queries~\cite{shrivastava2004medians}.
Our example pipeline in Figure~\ref{fig:sensor-node-internals}, shows how our solution complements acquisitional query processing.
We replace ad-hoc sensor reads with a sophisticated pipeline join which ensures time coherent result tuples.
Riahi et al. consider the efficient data acquisition from many sensors and optimize data gathering based on application requests~\cite{riahi2013Utility}. However, they do not discuss the time coherence of result tuples.
Barga et al. cope with variable latency and disorder in event streams correlation~\cite{barga2006consistent,barga2006coping}. They provide a service for central event correlation, which operates based on heartbeats and clock information carried in punctuation events.
In contrast to this service, SENSE avoids a complex central correlation and leverages edge computing capabilities to deliver time coherent data tuples.

\section{Conclusion}\label{sec:conclusion}
Upcoming IoT applications gather values from thousands of distributed sensors.
Currently, these applications do not detect clock offsets among sensor nodes and trust event timestamps without input validation.
We propose SENSE, the first system for sensor data acquisition that quantifies the time coherence of result tuples, provides synchronization independent coherence guarantees, and supports user-defined incoherence limits.
The core of SENSE are sensing loops, which gather sensor values sequentially from a set of sensors.
To ensure scalability and to enforce incoherence limits, SENSE dynamically splits and merges sensing loops with respect to time coherence and latency requirements.
Our results show that SENSE scales to thousands of sensor nodes and reliably optimizes the coherence estimate of tuples while keeping the coherence guarantee below a user-defined upper limit. Thereby, SENSE quickly adapts to changed network conditions and coherence requirements.

\bibliographystyle{ACM-Reference-Format}

\bibliography{references}  

\begin{appendix}

\pagebreak

\section*{Appendix}
\vspace{5mm}

\section{Derivation of the {\LARGE$\boldsymbol\alpha$}-Formula}\label{sec:derivationOfAlpha}
\vspace{3mm}

In this section, we mathematically derive Formula~\ref{eq:alpha}, which computes the optimal $\alpha$ on the loop node.
Our goal is to find the value for $\alpha$ which implies the minimum mean squared error between the request time $t$ of a tuple and the read times ($t_1,\dots,t_n$) of values ($v_1,\dots,v_n$) contained in the tuple.

Formula~\ref{eq:alphaMse} computes the mean squared error which depends on $\alpha$ and the join times $t_{now}^{(1)},\dots,t_{now}^{(N)}$ on sensor nodes.

\begin{align}
  \text{ERR}(\alpha, t_{now}^{(1)}, \dots, t_{now}^{(N)}) =  \frac{1}{N}\sum\limits_{i=1}^N \left(t - (t_{now}^{(i)} - \alpha) \right)^2\label{eq:alphaMse}
\end{align}~\\
We select $\argmin \limits_{\alpha} (\text{ERR}(\alpha, t_{now}^{(1)}, \dots, t_{now}^{(N)}))$ as $\alpha$ to minimize the error.
Because the actual join times at sensor nodes are unknown at the loop node, we replace them with estimated join times ($\mathbb{E}( t_{now}^{(1)}),
  \dots, \mathbb{E}(t_{now}^{(N)})$) in Formula~\ref{eq:alpaMse2}.
The expected value of $t_{now}^{(i)}$ is $l_s+\delta i$ because node $i$ receives the tuple $i$ hops after the loop start time $l_s$. The mean hop time is $\delta=\Delta/(N+1)$ and we know $\Delta$ from previous tuples.

\begin{align}
\text{ERR}\!\left(\!
  \alpha,\!\mathbb{E}\!\left(\!t_{now}^{(1)}\!\right)\!,\!
  \dots\!,\!\mathbb{E}\!\left(\!t_{now}^{(N)}\!\right)
\!\right)\!=\!\frac{1}{N}\!\sum\limits_{i=1}^N\!\left( t\!-\!( l_s\!+\!\delta i\!-\!\alpha)\!\right)^2
\label{eq:alpaMse2}
\end{align}~\\
We now define Formula~\ref{eq:alpaMse3} and insert it into Formula~\ref{eq:alpaMse2} which results in Formula~\ref{eq:alpaMse4}.

\begin{align}
\alpha=x+(l_s-t)\label{eq:alpaMse3}
\end{align}
\begin{align}
\text{ERR}\!\left(\!
  \alpha,\!\mathbb{E}\!\left(\!t_{now}^{(1)}\!\right)\!,\!
  \dots\!,\!\mathbb{E}\!\left(\!t_{now}^{(N)}\!\right)
\!\right)\!=\!\frac{1}{N}\!\sum\limits_{i=1}^N\!\left( x-\delta i \right)^2
\label{eq:alpaMse4}
\end{align}
~\\
We observe in Formula~\ref{eq:alpaMse5} that the second derivative of Formula~\ref{eq:alpaMse4} with respect to $x$ is positive.
Thus, the minimum of the function $\text{ERR}\left(\alpha, t_{now}^{(1)}, \dots, t_{now}^{(N)}\right)$ can be found at the simple zero of the first derivative with respect to $x$.

\begin{align}
\frac{\partial^2}{\partial^2 x} \text{ERR}\left(\alpha, t_{now}^{(1)}, \dots, t_{now}^{(N)}\right) = 2 > 0
\label{eq:alpaMse5}
\end{align}%
\begin{align}
  &~ \frac{\partial}{\partial x} \text{ERR}\left(
  \alpha, \mathbb{E}\left( t_{now}^{(1)}\right),
\dots, \mathbb{E}\left(t_{now}^{(N)}\right)\right)
  \nonumber\\
  &= 2N x  - 2\delta \sum \limits_{i=1}^{N} i
  = 2N x  - \delta N(N+1)\label{eq:alpaMse6}
\end{align}
~\\

We now set Formula~\ref{eq:alpaMse6} equal to $0$ and solve the equation for x which leads to Formula~\ref{eq:alpaMse7}. Inserting Formula~\ref{eq:alpaMse3} for $x$ leads to Formula~\ref{eq:alpaMse8} for the optimal $\alpha$.

\begin{align}
  x  &= \frac{\delta (N+1)}{2} = \frac{\Delta}{2}\label{eq:alpaMse7}\\
  \alpha  &= \frac{\delta (N+1)}{2}+(l_s-t) = \frac{\Delta}{2}+(l_s-t)\label{eq:alpaMse8}
\end{align}~\\

\section{Derivation of the {\LARGE$\boldsymbol\mu$}-Formula}\label{sec:derivationOfMu}
\vspace{3mm}

In this section, we derive an update rule for $\mu$, intending to change the coherence guarantee of the next tuple by $v$.
The direction of the change is determined by $\sign(D_{max}-C_g)$.\linebreak
$v$ directly translates to the step width $w$  introduced in Section~\ref{sec:sensor-node-algo} ($2w$=$v$).
We assume that $\Delta$ remains unchanged.
We aim to derive a function \textit{update} that satisfies the Condition~\ref{eq:apx:muUpdate}.

\begin{align}
  \mu_{new} &=  \textrm{update}\left(\mu, v\right)\qquad \textrm{s.th.}\label{eq:apx:muUpdate}\\
  C_{g} \left(\mu_{\text{new}}\right) &= C_{g}\left(\mu\right) + v\nonumber
\end{align}
~\\
\noindent In the remainder of this section, we first derive the optimal timestamp $t_{\textit{i;opt}}$ to be selected at an arbitrary node $i$ as a function of $\mu$.
We observe that the desired change $v$ in $C_g$ depends solely on the change in the distance between the largest and smallest age selected at a node.
We thus express $t_{\textit{i;opt}}$ as the optimal age $\alpha_{\textit{i;opt}}$ and compute the change $v$ relative to an update in $\mu$.
Finally, we discuss the relationship between $v$ and $w$ which leads to Formula~\ref{eq:mu} for the update of $\mu$.

\noindent Let $i\in\{1, \dots, N\}$ be an arbitrary but fixed node identifier. We aim to find the minimum of Cost Function\ref{eq:apx:cost}.

\begin{align}
  \text{cost}_{\mu, \alpha, t_{now}^{(i)}, t}\left(t_i\right) &= \abs (t_i-t) + \left ( t_{now}^{(i)} - t_i - \alpha\right)^2\mu\label{eq:apx:cost}
\end{align}
~\\
Function~\ref{eq:apx:cost} is convex because it is the sum of two convex functions~\cite{roberts1973convex}. It has exactly one global minimum for a fixed $\mu$.
However, Function~\ref{eq:apx:cost} is non-differentiable at $t_i$=$t$.
Therefore we must consider $t_i$=$t$ as a candidate for the global minimum in addition to two candidates $t_{i;1}$ and $t_{i;2}$ for the domains $A_1$:=$(-\infty,t)$ and $A_2$:=$(t, \infty)$.
The global minimum is located at the candidate which implies the smallest cost.

We compute $t_{i;1}$ and $t_{i;2}$ in Formula~\ref{eq:apx:opt12} which we determine based on the derivations of our cost function given in Formula~\ref{eq:derivate-cost} and~\ref{eq:derivate-cost2}.

\begin{align}
\frac{\partial}{\partial t_i} \textrm{cost}
\left(\cdot \right)
  &= \sign\left(t_i - t\right)
  + 2\mu \left(t_i - (t_{now}^{(i)}  - \alpha) \right) \nonumber\\
  &= \pm 1 \text{\hspace{12.37mm}}+ 2\mu \left(t_i - (t_{now}^{(i)}  - \alpha) \right) \label{eq:derivate-cost}\\
  \frac{\partial^2}{\partial^2 t_i} \textrm{cost} \left(\cdot \right)
  &= 2\mu  > 0 \label{eq:derivate-cost2}\\
  \frac{\partial}{\partial t_i} \textrm{cost}|_{A_{1,2}} &= 0
  \Leftrightarrow t_{i;1,2} = \frac{\mp 1}{2\mu}+ t_{now}^{(i)} - \alpha\label{eq:apx:opt12}
\end{align}
Plugging the values $t_{i;1,2}$ in the cost function reveals that the function's optimum lies between the optimum for the coherence guarantee $t_{now}^{(i)}$-$\alpha$ and the request timestamp $t$.
To condense the notation, we define the difference between $t$ and $t_{now}^{(i)}$-$\alpha$ as $r$ in Equation~\ref{eq:def-r}.
\begin{align}
r_i := t - \left(t_{now}^{(i)} - \alpha\right)\label{eq:def-r}
\end{align}
The evaluated cost function at each of the two candidates $t_{i;1,2}$ is given by Equation~\ref{eq:apx:optTEval}.
\begin{align}
  \text{cost}\left(t_{i;1,2}\right) &= \abs\left(\frac{\mp 1 + 2\mu r_i}{2\mu} \right) + \frac{1}{4\mu} \label{eq:apx:optTEval}
\end{align}
Considering which of the values $t_{i;1}$ or $t_{i;2}$ has lower cost we receive Equivalence~\ref{eq:apx:optTEval2}.
\begin{align}
\text{cost}\left(t_{i;1}\right)\!\substack{>\\<}\text{cost}\left(t_{i;1}\right)\!\Leftrightarrow \!\argmin\limits_{ t_{i;c}\in\{t_{i;1}, t_{i;2}\}}\!\!\!\!\left(\text{cost}(t_{i;c})\right)\substack{>\\<}t\!\Leftrightarrow\!0\substack{>\\<}r_i\label{eq:apx:optTEval2}
\end{align}
Using Equivalence~\ref{eq:apx:optTEval2} the selected candidate $t_{i;c}$ out of $t_{i;1,2}$ is given by Formula~\ref{eq:tic}.
\begin{align}
  t_{i;c} = \frac{\sign(r_i)}{2\mu}+ t_{now}^{(i)} - \alpha\label{eq:tic}
\end{align}
It remains to determine for which values of $\mu$ the minimum of the cost function is located at $t_{i;c}$ and for which values of $\mu$ it is located at $t$.
Therefore, we compare the evaluated cost function at $t_{i;c}$ and $t$ in Equivalence~\ref{eq:apx:muTTCEqui}.

\begin{align}
\text{cost}(t) &= r_i^2  \cdot \mu\nonumber \\
\text{cost}(t_{i;c}) &= \frac{-1 + 2\mu \abs(r_i)}{2\mu}  + \frac{1}{4\mu}\nonumber\\
  \text{cost}\left(t_{i;c}\right) &\leq  \textrm{cost}(t)
  \Leftrightarrow \mu \geq \frac{1}{2\abs(r_i)}\label{eq:apx:muTTCEqui}
\end{align}

We can now write the optimal value of $t_i$ as a function of $\mu$ in Definition~\ref{eq:tiStar}.
Definition~\ref{eq:aiStar} specifies the corresponding optimal age at node $i$.

\begin{align}
  t_{i;\text{opt}} \left(\mu\right):=
  \begin{cases}
    \frac{\sign(r_i)}{2\mu} + t_{now}^{(i)} - \alpha \quad& \mu \geq \frac{1}{2\abs(r_i)}\\
    t \quad& \textrm{otherwise.}
\end{cases}\label{eq:tiStar}\\
 \alpha_{i;\text{opt}} \left(\mu\right):=
  \begin{cases}
    \frac{-\sign(r_i)}{2\mu} + \alpha \hspace{8mm}\quad& \mu \geq \frac{1}{2\abs(r_i)}\\
    t_{now}-t \quad& \textrm{otherwise.}
  \end{cases}\label{eq:aiStar}
\end{align}~\\
The difference between all selected $\alpha_i$ values is equal to
$$ \frac{1}{2\mu} + \alpha - \left(\frac{-1}{2\mu} + \alpha \right) = \frac1\mu.$$
This leads to Formula\ref{eq:apx:cg} for the coherence guarantee $C_g$.

\begin{align}
  C_g = \min\left(\Delta + \frac1{\mu}, 2\Delta\right) \label{eq:apx:cg}
\end{align}~\\
If $\mu>\tfrac{1}{r_1}$, we achieve the desired change of $v$ in $C_g$ by updating $\mu$ according to Formula~\ref{eq:apx:vupdate}.
Otherwise, the old value of $\mu$ is already optimal.

\begin{align}
  \frac1{\mu_{new}}  = \frac1{\mu}+v
  \Leftrightarrow  \mu_{new}  = \frac\mu{1 + \mu v}\label{eq:apx:vupdate}
\end{align}~\\
By now we derived the $\mu$ update rule depending on $v$.
We now discuss the relation between $v$ and $w$ and derive the $\mu$ update rule depending on $w$ in Equation~\ref{eq:apx:wupdate}.
In case $D_{max}$ was met, we target to improve the coherence estimate $C_e$ for the next tuple at the cost of increasing $C_g$ closer to $D_{max}$.
Otherwise, $C_g$ needs to be reduced. Thus, the direction of the $\mu$ update is given by $\sign(D_{max}-C_g)$.
The optimal age $\alpha_{i;opt}$  needs to be reduced or increased depending on the position $i$ in the pipeline. Whether we need to increase or decrease the age $\alpha_{i;opt}$ is determined by $\sign(r_i)$ in the first line of Equation~\ref{eq:apx:wupdate}.
In the general update rule, which is independent of $i$, $\sign(r_i)$ cancels out (Line 2 in~Equation~\ref{eq:apx:wupdate}).

\begin{align}
\alpha_{i;opt}\left(\mu_{new}\right) &= \alpha_{i;opt}\left(\mu\right) + \sign(r_i)\sign(D_{max} - C_g)w\nonumber\\
\Leftrightarrow \frac{1}{\mu_{new}} &= \frac{1}{\mu} -2\sign(D_{max} - C_g)w\nonumber\\
\Leftrightarrow \mu_{new} &= \frac{\mu}{2\sign(D_{max} - C_g)w\mu +1}\label{eq:apx:wupdate}
\end{align}~\\

Comparing Equation~\ref{eq:apx:vupdate} with Equation~\ref{eq:apx:wupdate} yields that \mbox{$v$=$2w$}.
Line 3 of~Equation~\ref{eq:apx:wupdate} is the final general $\mu$ update rule.

\vspace{3mm}
\section{Derivation of the {\LARGE$\boldsymbol\mu$}-Initialization rule}\label{sec:InitializationOfMu}
\vspace{3mm}

Even though the method of selecting the update step width $w$ is very efficient and ensures fast convergence, it is reasonable to initialize $\mu$ taking into consideration the targeted $C_g$-$C_e$-tradeoff.
In the Appendix~\ref{sec:derivationOfMu}, we derived
$C_g = \Delta + \tfrac1\mu$
in Equation~\ref{eq:apx:cg} for $\mu > \tfrac1{r_1}$.
After estimating the round trip time $\Delta$, we can use $D_{max}$ for initializing $\mu$ to optimize the coherence estimate while maintaining the upper bound of the guarantee.
We aim to set $C_g = D_{max} - \Delta$ which is achieved by initializing $\mu$ according to Equation~\ref{eq:apx:muInitRule}.

\begin{align}
  \mu &= \frac{1}{D_{max} - 2\Delta} \label{eq:apx:muInitRule}
\end{align}~\\

\section{Clock Specifications}\label{sec:clocks-for-clock-drift}

\vspace{4mm}
\begin{table}[t]
\setlength{\tabcolsep}{3pt}
\centering
\begin{tabular}{|l|r|c|r|}
\hline
\textbf{Clock}            & \multicolumn{1}{l|}{\textbf{PPM}} & \multicolumn{1}{l|}{\textbf{Price}} & \multicolumn{1}{l|}{\textbf{Frequency}} \\ \hline
Raspberry Pi System Clock~\cite{bergsma2013howaccurate} & 40                                & free                                & 1 MHz                                   \\ \hline
PCF2127 Real Time Clock~\cite{nxp2014PCF2127}   & 3                                 & 14\$                          & 32.768 kHz                              \\ \hline
449-LFTVXO076344CUTT~\cite{iqd2017PCF2127}      & 0.1                               & 23\$ & 19.2 MHz                                \\ \hline
\end{tabular}
\vspace{3mm}
\caption{Specifications of different clocks at 30°C}\label{tab:sensor-specs}
\end{table}

In Figure~\ref{fig:clock-drifts} of Section~\ref{sec:sources-of-incoherence}, we presented the clock drifts of different clocks.
The figure shows the clock drifts of the selected clocks at 30°C.
Table~\ref{tab:sensor-specs} summarizes the clock specifications of the selected clocks.
The \textit{parts per million} (PPM) provide an upper bound to the amount of additional or missed oscillations.
For example, the Raspberry Pi system clock has a frequency of 1MHz (i.e. $10^6$ oscillations per second) and drifts with 40PPM.
Thus it drifts by $\pm40/10^6$ seconds per second which is a drift of $\pm$0.144 seconds per hour.

We obtained the data for Figure~\ref{fig:clock-drifts} by simulating the clock on the level of its oscillator.
The expected amount of missed  oscillations of a specific clock is assumed to be uniformly distributed within the ranges the manufacturer provided. This value -- in the following referred to as $p$ -- is sampled once per clock instance.
The oscillator is modelled as a Bernoulli process. Thus, each time a timestamp $t_{i+1}$ is requested from a clock instance, we sample a Binomial distribution with probability $1+p$ and the amount of trials equal to the time that passed since the last request $(t_{i+1} - t_{i})$ in seconds multiplied by the clock model's frequency.

\section{Trilateration Calculations}\label{sec:trilateration-calc}
\vspace{3mm}

In this section, we summarize the calculations which lead to the example provided in Section~\ref{sec:intro-example} and Figure~\ref{fig:intro-example}.

We first define the position of sensors as
$s_1(100,200)$,\linebreak
$s_2(7000,1000)$, and
$s_3(4200,4000)$.
The real location of the signal source in our example is $S(3456,1234)=(x,y)$.

\begin{figure}[t]
	\centering
	\includegraphics[width=0.7\linewidth]{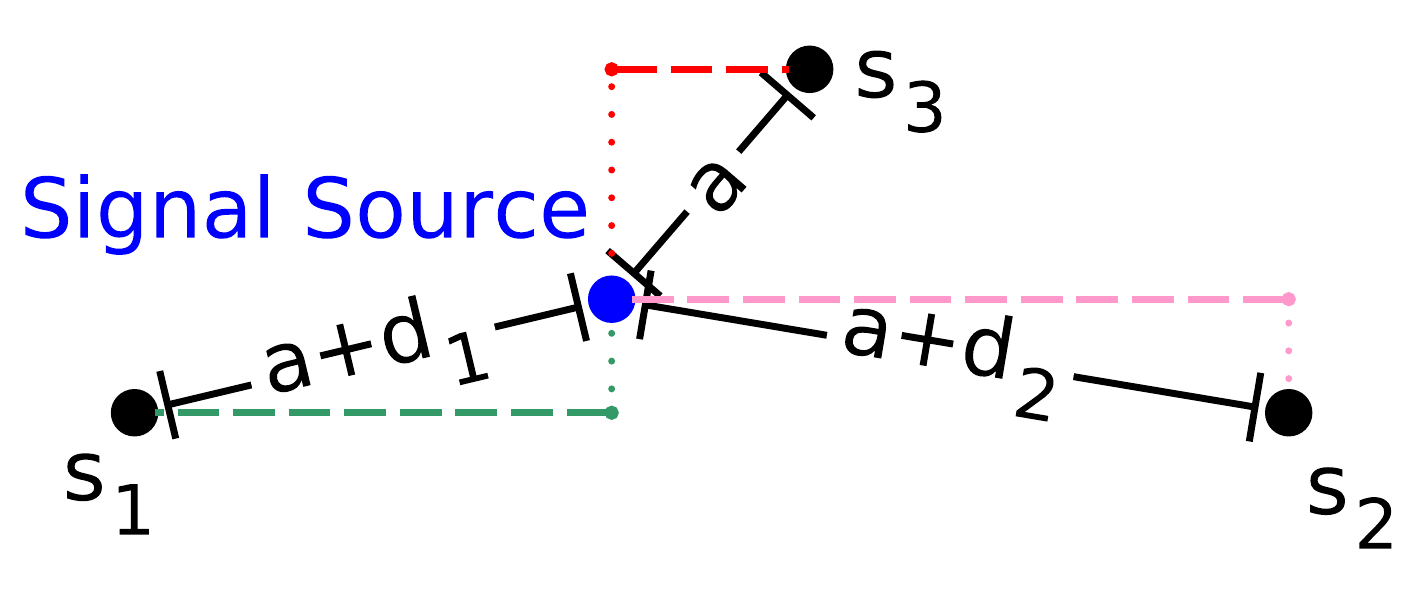}
	\caption{Sketch of the distances between the signal source and the sensors.}\label{fig:sensor-sketch}
\end{figure}

We measure the arrival times of the signal at our three sensors.
$t_1$, $t_2$, and $t_3$ are the times when the signal arrives at $s_1$, $s_2$, and $s_3$.
The signal arrives first at the closest sensor which is $s_3$.
Then, the signal arrives $t_1-t_3$ later at $s_1$ and $t_2-t_3$ later at $s_2$.
We can convert these time frames to distances by multiplying them with the signal dissemination speed. In our example this is sonic speed which is 343m/s. We name the resulting distances $d_1$ and $d_2$. In addition we name the unknown distance between $s_3$ and the signal source $a$. Figure~\ref{fig:sensor-sketch} provides a sketch of the distances between the signal source and the sensors.
With the Pythagorean Theorem, we derive the Equation System~\ref{eq:sensors-eq-1}.

\begin{equation} \label{eq:sensors-eq-1}
\definecolor{s1triangle}{HTML}{339966}
\definecolor{s2triangle}{HTML}{FF99CC}
\begin{split}
a^2 & =\udash{red}{(4200 -- x)}^2+\udot{red}{(4000 -- y)}^2 \\
(a+d_1)^2 & =\udash{s1triangle}{(100 -- x)}^2+\udot{s1triangle}{(200 -- y)}^2 \\
(a+d_2)^2 & =\udash{s2triangle}{(7000 -- x)}^2+\udot{s2triangle}{(1000 -- y)}^2
\end{split}
\end{equation}

With perfectly synchronized sensor node clocks, we measure $d_1\approx647.37$ and $d_2\approx687.40$. We can now solve the equation system for $a$, $x$, and $y$ to receive the position of the signal source $x=3456$, $y=1234$, and $a=2864.31$.
Any localization failure can be computed by adjusting $d_1$ and $d_2$ with respect to wrongly measured signal arrival times.

The coherence guarantee $C_g$ quantifies the uncertainty about the precision of signal arrival times. Given $l_s$, $l_e$, $\alpha_{min}$, and $\alpha_{max}$, we can compute a time period in which the signal arrived at a sensor.
The earliest possible arrival time of the signal is $l_s-\alpha_{max}$.
The latest possible arrival time is $l_e-\alpha_{min}$.
The time frame between the earliest and the latest possible time is $C_g$.
To get the guaranteed location precision (orange area in Figure~\ref{fig:intro-example}), we calculate the earliest and latest possible arrival times for all three sensors.
We then calculate six localizations of the signal source which result from combinations of earliest and latest possible arrival times.
These are all combinations excluding the two combinations which include all earliest and all latest arrival times.
The resulting positions are the corners of the guaranteed detection area.
One can observe that changing any detection time from its minimum to its maximum results in a linear movement of the computed location from one corner to another corner of the guarantee area.
Thus, the real location of the signal source must be inside this area.

We compute the area of the precision estimate analogue to the area of the precision guarantee with $t_{min}$ being the earliest arrival time and $t_{max}$ being the latest arrival time.

\vspace{3mm}
\section{Periodic Scheduling Results}\label{app:periodic_scheduling}
\vspace{3mm}

\begin{figure}[t]
	\centering
	\includegraphics[]{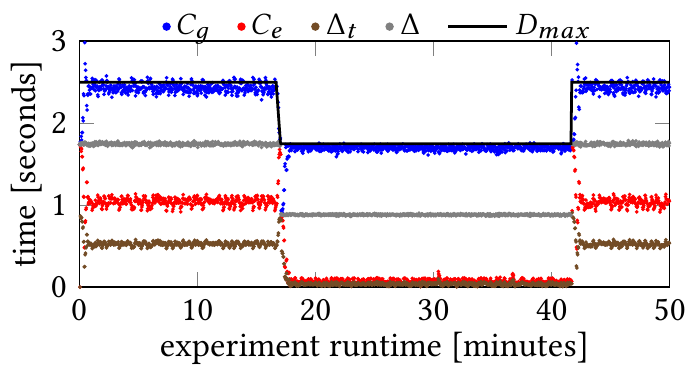}
	\caption{Evolution of coherence guarantees and estimates for changing incoherence limits ($\boldsymbol D_{max}$).\\ Experiment results with periodic scheduling.}\label{fig:triple-plot-appendix}
\end{figure}

In Figure~\ref{fig:triple-plot-appendix}, we show the experiment results for \textit{periodic scheduling}.
The plot looks identical to the result for \textit{Schedule Next Read} show in Figure~\ref{fig:triplePlot-SNR} in Section~\ref{sec:exp-opt-coherence}.
The only exception is the read time deviation $\Delta_t$ which is smaller for \textit{Schedule Next Read} because we can schedule sensor reads precisely at the requested times.

\vspace{3mm}
\section{Additional Optimizations}\label{sec:opt-at-node}
\vspace{3mm}

We apply an additional optimization on sensor nodes which utilizes our knowledge about read times on previous nodes in the loop.
More precisely, at any sensor node, we know the current values of $\alpha_{min}$, $\alpha_{max}$, $t_{min}$, and $t_{max}$.
Since $C_e$ and $C_g$ do not change, if we change neither $\alpha_{min}$, $\alpha_{max}$, $t_{min}$, nor $t_{max}$, we can select measurements from the history buffer which do not change these values without a negative impact on $C_e$ or $C_g$ improving $\Delta_t$.
In Formula~\ref{eq:optFunc}, we showed the combined cost function for deviations from $C_e$ and $C_g$.
The underlying individual cost functions are Formula~\ref{eq:cost-ce} for $C_e$ and Formula~\ref{eq:cost-cg} for $C_g$.

\begin{align}
\text{cost}_{C_e}(t,t_i)&=\abs(t_i-t)\label{eq:cost-ce}\\
\text{cost}_{C_g}(t_{now},t_i,\alpha)&=(t_{now}-t_i-\alpha)^2\label{eq:cost-cg}
\end{align}~\\
We extend these cost functions to express that selecting a value with $t_{now}-t_i = \alpha_i \in [\alpha_{min},\alpha_{max}]$ implies no costs  with respect to $C_g$ and that selecting a value with $t_i \in [t_{min},t_{max}]$  implies no costs with respect to $C_e$.

This optimization allows for utilizing failures of previous nodes to select a value which is as close as possible to $t$ if this does neither increase $C_e$ nor $C_g$. Moreover, the extended cost functions prevent increasing $C_e$ or $C_g$ if there exists a value in the buffer which does not increase $C_e$ or $C_g$.

\end{appendix}

\end{document}